\documentclass[12pt,superscriptaddress,notitlepage,nofootinbib]{revtex4-1}
\usepackage[utf8x]{inputenc}
\usepackage[T1]{fontenc}
\usepackage{amsmath}
\usepackage{amssymb}
\usepackage{amscd}
\usepackage{amsfonts}
\usepackage{amstext}
\usepackage{amsthm}
\usepackage{graphicx}
\usepackage{color}

\def\CC {{\mathbb C}}
\def\II {{\mathbb I}}
\def\NN {{\mathbb N}}    
\def\RR {{\mathbb R}}     
     
\def\det {\mathrm{det}\, }
\def\tr  {\mathrm{tr}\, }
\def\Im {\mathrm{Im}\, }
\def\Re {\mathrm{Re}\, }

\begin{document}
\title{Trace formulas for general Hermitian matrices:
  Unitary scattering approach and periodic orbits on an associated graph}
\author{Sven~Gnutzmann}
\email{sven.gnutzmann@nottingham.ac.uk}
\affiliation{School of Mathematical Sciences, University of Nottingham, University Park, Nottingham NG7 2RD, UK}
\author{Uzy~Smilansky}
\email{uzy.smilansky@weizmann.ac.il}
\affiliation{Department of Physics of Complex Systems, Weizmann Institute of Science, Rehovot 76100, Israel}

\begin{abstract}
  Two trace formulas for the spectra of arbitrary Hermitian matrices are
  derived by
  transforming the given Hermitian matrix $H$ to a unitary analogue.
  In the first type the unitary matrix is $e^{i(\lambda\II - H)}$ where
  $\lambda$ is
  the spectral parameter. The new feature  is that the spectral parameter
  appears in the final
  form as an argument of  Eulerian polynomials -- thus connecting the periodic
  orbits to
  combinatorial objects in a novel way. To obtain the second type, one
  expresses the input
  in terms of a unitary scattering matrix in a larger Hilbert space.
  One of the surprising 
  features here is that the
  locations and radii of the spectral discs of Gershgorin's theorem
  appear naturally as the pole parameters
  of the scattering matrix. 
  Both formulas are discussed and possible applications are outlined.    
\end{abstract}

\maketitle
\section{Introduction}
\label{intro}

Trace formula is a generic name for relations which connect between spectral
information and geometric or dynamical information pertaining to the same
operator and its domain. Since it was first introduced by
A.~Selberg \cite{Selberg}, it has been one of the main tools in many fields of
research, ranging between number theory, spectral geometry,
graphs (combinatorial and quantum) and the semi-classical theory of
integrable and chaotic area preserving dynamical systems
\cite{Gutzwiller1970,Gutzwiller1971,Gutzwillerbook,Balianbloch,Berrytabor,Ozorio,qsoc}.
Typically, the spectral information is expressed in terms of the spectral
density or the spectral counting function.  The geometrical information
resides in the properties of periodic structures or orbits, such as closed
walks on connected vertices in a graph,  periodic classical trajectories,
periodic geodesics on closed surfaces, \emph{etc}.

A simple example arises in the study of the spectra finite of
Hermitian matrices. The spectral density of a $N \times N$
Hermitian matrix $H$ with spectrum denoted by $\{\lambda_j\}_{j=1}^N$
can be expressed in terms of the resolvent (Green function)
$G(z) = (z\II -H)^{-1}$
\begin{equation}
  \rho (\lambda) =\sum_{j=1}^N  \delta({   \lambda}-\lambda_j) =
  \frac{1}{\pi}\lim_{\epsilon\rightarrow 0} \Im  \left[\tr G(\lambda+i\epsilon)\right].
  \label{resolvent}
\end{equation}
{  The Green function $G(z)$ is a meromorphic function on
  the complex plane $z \in \mathbb{C}$ with poles
  at the eigenvalues of $H$. 
  While the expansion
  $\tr G(z)=\sum_{n=0}^{\infty} \frac{\tr H^n}{z^{n+1}}$ only converges
  absolutely for $z$ outside the spectral radius, one may substitute this
  expansion in \eqref{resolvent} and take $\epsilon \to
  0$ if one considers this as a distributional identity where the series
  converges subject to integration with respect to an appropriate
  set of test functions.
  In this sense one obtains
}
\begin{equation}
  \rho (\lambda) =\frac{1}{\pi}\lim_{\epsilon\rightarrow 0}
  \Im  \left[\sum_{n=0}^{\infty} \frac{\tr H^n}{(\lambda+i\epsilon)^{n+1}}\right].
  \label{resolvent1}
\end{equation}
Thus, the spectral density is expressed in terms of the geometrical
information embedded in $\tr H^n$. The connection to ``geometry''
comes by recalling that
\begin{equation}
  \tr H^n = \sum _{\{1\le i_k\le N \}} H_{i_1,i_n} H_{i_n,i_{n-1}}\cdots H_{i_2,i_1}
  \label{trace_periodic_orbits}
\end{equation}
which can be viewed as a sum over weighted periodic walks
on a connected
graph on $N$ vertices (with self loops but no {  multiple} edges),
where the weight on the edge  ${i,j}$ is the matrix element $H_{i,j}$.
When $H$ is e.g., an adjacency matrix of a combinatorial graph,
with $H_{i,j}\in [0,1])$,  $\tr H^n$
counts the number of periodic walks with period $n$ on the graphs.
This connection was used by E.~Wigner \cite{wigner}
in deriving the semi-circle law by counting the mean number of
periodic walks on random graphs.

The above example also illustrates an important characteristic of
trace formulas: when they are written as equalities, they  are formal,
and cannot be used as numerical ones. Rather, they express relations
whose contents can be elucidated by analytical continuation,
or by application to appropriate test functions. 

In the present work we derive and study two types of trace formulas,
which, to the best of our knowledge were not previously formulated.
Both of them express the spectral density or the spectral counting
function of a Hermitian matrix $H$
in terms of (weighted) periodic orbits on an underlying graph
in the sense explained above.
The common feature of both approaches is that the spectral data of
the original matrix $H$ is stored in a unitary matrix $S(\lambda)$ such that
$\lambda=\lambda_j$ is an eigenvalue of $H$ if and only if the
unitary matrix $S(\lambda)$
has a stationary eigenvector (that is an eigenvector $\mathbf{x}$ that
satisfies $S(\lambda)\mathbf{x}=\mathbf{x}$).
The two kinds of trace formulas that we present are derived from different
choices of the matrix $S(\lambda)$.

In Sec.~\ref{tf1} we present the first approach where
$S(\lambda)$ has the same
dimension $N$ as $H$ and is constructed from the ``evolution operator''
$e^{- i H t}$ for a unit of time $t$
(without loss of generality we will set $t =1$). 
In this case the trace formula will be expressed
in two different
forms that involve the traces $\tr H^n$ and thus the same periodic orbits
and weights on an underlying graph as described in
\eqref{trace_periodic_orbits}.
In the final expression, the Polylogarithmic  functions of
negative index (which may be expressed in terms of Eulerian polynomials) play a crucial role. 

In Sec.~\ref{tf2} we present a rather different approach.
There, we construct a unitary matrix $S(\lambda)$ that
may be viewed as  an
evolution operator of a discrete time quantum walk
on the directed edges of the underlying
simple graph without the self-loops present in the first approach. The
dimension of
the $S$ matrix is larger than $N$ and equals the number of non
vanishing off-diagonal  matrix elements (if $H_{vw}=H_{wv}^*\neq 0$ and
$w\neq v$ both matrix elements are counted).
It enables writing a trace formula in yet another form,
albeit the input to the effective weights and the phases carried
by the paths, are different than the ones used to compute
$\tr H^n$ in the first approach. This second approach is a generalization
of previous work \cite{uzy} where
a similar trace formula has been derived for discrete graph Laplacians.
In spectral graph theory there is a long history of expressing
the spectra of (weighted) graph Laplacians in terms of walks or periodic orbits
on the
directed edges. 
Starting from the Ihara $\zeta$-function \cite{ihara1,ihara2} these
use a number of {  determinantal}
equalities that have been developed by various
authors
\cite{traceformula_regI,traceformula_regII,hashimoto1,hashimoto_hori,hashimoto2,hashimoto3,bass,bartholdi,kotani-sunada,nalini}.
Our approach leads to a {  determinantal}
equality that is, to the best
of our
knowledge new and belongs in the same context. A main difference between
our approach and the ones found in the literature is its generality
as one may start from an arbitrary complex Hermitian matrix while the most
general equalities available in the literature are for graph
Laplacians
\cite{nalini} (a subset of real symmetric matrices).
In the Ihara trace formula and many of its generalizations the sum over
periodic orbits is reduced to non-backscattering orbits.
This simplification does not apply to our approach however.

In Sec.~\ref{conclusions} we conclude this paper with a general outlook.

{  Note: in order to make a clear distinction between
  quantities related to the two trace formulas, we shall distinguish
  them by a suffix $I$ or $II$. }

\section{Approach I: a trace formula based on the time evolution operator}
\label{tf1}

Let $H$ be a Hermitian matrix of dimension $N$,
with spectrum $\{\lambda_i\}_{i=1}^N$ consisting of  the roots of the
characteristic polynomial
\begin{equation}
  \zeta_H(\lambda)=\det (\lambda\II - H).
\end{equation}
We write the Heaviside step function as $\theta(x)$ and denote by 
\begin{equation}
  \mathcal{N}(\lambda)=\sum_{j=1}^N \theta(\lambda-\lambda_j)
\end{equation}
the number of eigenvalues (zeros of the characteristic polynomial with multiplicities) 
which are smaller or equal $\lambda$.
We will refer to $\mathcal{N}({  \lambda})$ as the
(spectral) counting function and its (formal) derivative
\begin{equation}
  \rho(\lambda)= \sum_{j=1}^N \delta (\lambda- \lambda_j)
\end{equation}
as the density of states.

In this section we shall present two variants of a
trace formula which express $\mathcal{N}(\lambda)$
%for
in terms of the traces of
powers of $H$ in a different way than given
by \eqref{resolvent1} in the introduction
Sec~\ref{intro}.
For this we consider the unitary $N \times N$ matrix
\begin{equation}
  S_I(\lambda)= e^{i(\lambda\II -H)}
\end{equation}
such that $S_I(0)= e^{- i H}$ is equal to the time evolution operator
$e^{- i H t}$ for a unit of time $t$ where the units have
been chosen such that $t=1$.
Without loss of generality we assume that the spectrum is
normalized and shifted such that it is in the real interval $(-\pi,\pi)$,
and ordered monotonically $\lambda_i\le \lambda_{i+1}$.
This ensures that there is a one-to-one and order-preserving
correspondence between
the eigenvalues $\{e^{-i \lambda_j}\}$ of $S_I(0)$ on the unit circle and the eigenvalues
$\{\lambda_j\}$ of $H$.
Moreover, strict positivity of
the gap
\begin{equation}
  \delta:=\pi - \mathrm{max}(|\lambda_i|)=
  \mathrm{min}(\pi-\lambda_N, \lambda_1+\pi)>0
  \label{gap}
\end{equation}
will be assumed in order to ensure convergence
of the trace formulas. This restriction can always be met by a trivial
rescaling $ H \mapsto c H$ with some $c \in \RR$.  
We also introduce the secular function
\begin{equation}
  \zeta_I(\lambda)= \det \left(\II-S_I(\lambda) \right)
\end{equation}
which is a periodic function $\zeta_I(\lambda+2\pi)=\zeta_I(\lambda)$
with $N$ zeros in the interval $\lambda \in (-\pi,\pi)$ at the eigenvalues
$\lambda = \lambda_j$ of $H$.

\subsection{The first variant of the trace formula based on $S_I(\lambda)$}

Restricting $\lambda \in (-\pi, \pi)$ the first variant of the trace formula
is given by
\begin{subequations}
  \label{tracepoly}
  \begin{align}  
    \mathcal{N}_I(\lambda) =
    &
      \overline{\mathcal{N}}_I(\lambda) +
      \mathcal{N}_{I}^{(\mathrm{osc})}(\lambda)\\
    \label{tracepoly_weyl}
    \overline{\mathcal{N}}_I(\lambda)=
    & 
      \frac{ N \lambda - \tr H}{2\pi}+\frac{N}{2}\\
    \label{tracepoly_osc}
    \mathcal{N}_I^{(\mathrm{osc})}(\lambda)=
    &
      \frac{1}{\pi} \lim_{\epsilon \rightarrow  0^+}
      \Im \left \{ \sum_{n=1}^{\infty} \sum_{s=0}^{\infty}  e^{-i\frac{\pi}{2}s}
      \ \frac {\tr H^s}{s !}\ n^{s-1}\ e^{(i \lambda -\epsilon)n}
      \right \}\nonumber\\
    =
    &
      \frac{1}{\pi} \lim_{\epsilon \rightarrow  0^+}
      \Im \left \{  \sum_{s=0}^{\infty}   e^{-i\frac{\pi}{2}s}
      \ \frac {\tr H^s}{s !}\sum_{n=1}^{\infty} n^{s-1}\ e^{(i \lambda -\epsilon)n}
      \right \}  
      \ .
  \end{align}
\end{subequations}
The corresponding expression for the density of states reads
\begin{subequations} \label{trace_density}
  \begin{align}
    \rho_I(\lambda)=
    &
      \overline{\rho}_I(\lambda) + \rho_I^{(\mathrm{osc})}\\
    \label{trace_density_weyl}
    \overline{\rho}_I(\lambda)=
    &
      \frac{  N }{2\pi}\\
    \label{trace_density_osc}
    \rho_I^{(\mathrm{osc})}=
    &
      \frac{1}{\pi} \lim_{\epsilon \rightarrow  0^+}
      \Re \left \{  \sum_{s=0}^{\infty}  \  e^{-i\frac{\pi}{2}s}
      \ \frac {\tr H^s}{s !}\sum_{n=1}^{\infty} n^s\ e^{(i \lambda -\epsilon)n}
      \right \}  
      \ .
  \end{align}
\end{subequations}
We will refer to $\overline{\mathcal{N}}_I({ \lambda})$ and
$\overline{\rho}_I(\lambda)$ as the smooth part and  to
$\mathcal{N}_I^{(\mathrm{osc})}(\lambda)$ and 
$\rho_I^{(\mathrm{osc})}$ as the oscillating parts.
One should note that we have defined
$\mathcal{N}_I(\lambda)$ and $\rho_I(\lambda)$ 
only for $|\lambda| < \pi$
while $\mathcal{N}(\lambda)$
and $\rho(\lambda)$ are defined on the real line.
If one uses the given expressions for the trace formulas
one finds that the spectrum is repeated periodically
such that $\rho_I(\lambda + 2\pi)=\rho_I(\lambda)$,
$\overline{\mathcal{N}}_I(\lambda+2\pi)= N+
\overline{\mathcal{N}}_I(\lambda)$ and
$\mathcal{N}_I^{(\mathrm{osc})}(\lambda+2\pi)
=\mathcal{N}_I^{(\mathrm{osc})}(\lambda)$
so outside the interval $|\lambda|<\pi$
one generally has $\mathcal{N}_I(\lambda) \neq
\mathcal{N}(\lambda)$ and $\rho_I(\lambda)\neq \rho(\lambda)$.
This periodic behaviour makes this approach fundamentally different from the
second approach that we will lay out in Sec.~\ref{tf2} where no such
periodicity will appear.

The trace formula \eqref{tracepoly} can be derived {  from the secular equation}
$\zeta_I (\lambda) =0$.
Using Cauchy's theorem,
the identity $\log \det (\II-S)=\tr \log (\II-S)$ and expanding $e^{-iH}$ in a
Taylor series gives \eqref{tracepoly}.
Note that while the double sum in the limit converges for any $\epsilon > 0$
it \emph{does not converge absolutely}
for arbitrarily small $\epsilon >0$ as 
\begin{equation}
  \sum_{n=1}^{\infty} \sum_{s=0}^{\infty}
  \left|
    e^{-i\frac{\pi}{2}s}
    \ \frac {\tr H^s}{s !}\ n^{s-1}\ e^{(i \lambda -\epsilon)n}
  \right| =\sum_{n=1}^\infty \sum_{j=1}^N \frac{e^{n \left(|\lambda_j|-\epsilon\right)}}{n}
\end{equation}
which converges only if $\epsilon > \mathrm{max}_{j=1}^N (|\lambda_j|)$
{  (thus, for any given matrix $H$ with a  spectrum
  that satisfies the assumed restrictions one may choose
  $\epsilon = \pi$ to ensure convergence.)}
%  uniform convergence over all allowed matrices $H$)}.
Nonetheless we formally interchange the two summations in the last line of
\eqref{tracepoly_osc} in order to
arrive at an expression where $\mathrm{tr}\ H^s$ (and thus
{  the periodic orbits})
can be attributed a specific weight. 
Note that almost all interesting trace formulas express distributional
identities. 
This applies to the present case as well and the interchange
of summation just implies a suitable restriction to test functions that render
the double summation absolutely convergent. In this case this is achieved
by
considering test functions $f(\lambda)$ defined for $\lambda\in (-\pi,\pi)$
by a Fourier series $f(\lambda)= \sum_{n=-\infty}^\infty f_n e^{i\lambda n}$
with $\lim \mathrm{sup}\ e^{\pi n} |f_n| < C$ for some constant $C$.
This condition implies that the test functions are analytic on the real
axis\footnote{In general, a periodic function that is analytic on the real
  line has Fourier coefficients that decay exponentially with \emph{some}
  exponent. Our requirement that the exponent is at least $\pi$ implies that
  we are dealing with very smooth analytic functions.}.
Trace formulas are often used as formal devices far beyond the rigorous
applicability. Indeed we will show later (in
Sec.~\ref{sec:application_spectral_averages}) that this trace formula gives
correct results even when formally applied to test functions that do not
obey the stated rigorous restrictions. {  For practical
  purposes one
  can achieve numerical convergence in the trace formula
  \eqref{tracepoly} for
  the spectral counting function by introducing appropriate
  $\epsilon$-dependent cut-offs for the double sum, see App.~\ref{appendix_convergence}.}

An equivalent derivation starts with the
expression
\begin{align}
  \theta(\lambda-\lambda_j)=
  & \frac{\lambda-\lambda_j}{2\pi}+
    \frac{1}{2}- \lim_{\epsilon \to 0} \mathrm{Im}\
    \log \left( 1- e^{i(\lambda-\lambda_j)-\epsilon}\right)\nonumber \\
  =& \frac{\lambda-\lambda_j}{2\pi}+ \frac{1}{2}+ \lim_{\epsilon \to 0} \mathrm{Im}
     \sum_{n=1}^\infty \frac{e^{in(\lambda-\lambda_j)- n
     \epsilon)}}{n}\nonumber \\
  =& \frac{\lambda-\lambda_j}{2\pi}+ \frac{1}{2}+ \lim_{\epsilon \to 0} \mathrm{Im}
     \sum_{n=1}^\infty \sum_{s=0}^\infty e^{-i \frac{\pi}{2} s} \frac{\lambda_j^s}{s!}
     n^{s-1}e^{(i\lambda-\epsilon)n}
\end{align}
which is valid  for $\lambda\in (-\pi, \pi)$.
Summing this identity 
over the $N$  eigenvalues $\{\lambda_j\}$ then gives the trace formula
\eqref{tracepoly}.

\subsection{The second variant of the trace formula based on $S_I (\lambda)$:
  Polylogarithms and Eulerian polynomials }

The second variant  of the trace formula is obtained
by noting that the  sum over $n$ in the trace formula \eqref{tracepoly} can be
expressed in terms of the Polylogarithm functions
\begin{equation}
  \mathrm{Li}_{-s}\left(z\right):=\sum_{n=1}^\infty n^s z^n.
\end{equation}
Performing the sum over $n$ in \eqref{tracepoly}
one then obtains the trace formula
\begin{subequations}
  \label{tracepoly1}
  \begin{align}
    \mathcal{N}_I(\lambda) =
    &
      \overline{\mathcal{N}}_I(\lambda) +
      \mathcal{N}_I^{(\mathrm{osc})}(\lambda)\\
    \label{tracepoly1_weyl}
    \overline{\mathcal{N}}_I(\lambda)=
    &
      \frac{  N \lambda - \tr H}{2\pi}+\frac{N}{2}\\
    \label{tracepoly1_osc}
    \mathcal{N}^{(\mathrm{osc})}_I(\lambda)=
    &
      \frac{1}{\pi} \lim_{\epsilon \rightarrow  0^+}
      \Im \left \{ \sum_{s=0}^{\infty} \ e^{-i\frac{\pi}{2}s}
      \ \frac {\tr H^s}{s !}\ \mathrm{Li}_{-s+1}\left(e^{(i \lambda -\epsilon)}\right)
      \right \}
  \end{align}
\end{subequations}
for the spectral counting function, and
\begin{subequations}
  \label{trace_density1}
  \begin{align}
    \rho_I(\lambda) =
    &
      \overline{\rho}_I(\lambda)+
      \rho^{(\mathrm{osc})}_I(\lambda)
    \\
    \label{trace_density1_weyl}
    \overline{\rho}_I(\lambda)=
    &
      \frac{N}{2\pi}\\
    \label{trace_density1_osc}
    \rho_I^{(\mathrm{osc})}(\lambda)=
    &
      \frac{1}{\pi} \lim_{\epsilon\rightarrow 0^+} \Re
      \left \{ \sum_{s=0}^{\infty} \ e^{-i\frac{\pi}{2}s}
      \ \frac {\tr H^s}{s !}\ \mathrm{Li}_{-s}\left(e^{(i \lambda -\epsilon)}\right)
      \right \}
  \end{align}
\end{subequations}
for the density of states.
The relevant Polylogarithms are explicitly given by
\begin{align}
  \mathrm{Li}_{0}(z)&= \frac{z}{1-z}\\
  \mathrm{Li}_{1}(z)&=-\log (1-z)
\end{align}
and the recursion
\begin{equation}
  \mathrm{Li}_{-s}(z)=z \frac{\partial \mathrm{Li}_{1-s}(z)}{\partial z}
  = \left( z \frac{d}{dz}\right)^s\frac{z}{1-z}
  \label{Li_recursion}
\end{equation}
for $s>0$.
This implies that the 
Polylogarithms with negative index are rational functions 
\begin{align}
  \label{Eulpoldef}
  \mathrm{ Li}_{-s}(z) = \frac{z}{(1-z)^{s+1}}
  A_s(z)
\end{align}
where $A_s(z)$ are known to be 
the Eulerian polynomials \cite{Eulerian} (not to be confused with the Euler polynomials)
which may be written as
\begin{equation}
  A_s(z)=\sum_{k=0}^{s-1}  A(s,k)z^k
  \label{Eulerpolynomial}
\end{equation}
in terms of the Eulerian numbers
\begin{equation}
  \label{Eulnumdef}
  A(s,k) =\sum_{m=0}^{k} (-1)^m
  \begin{pmatrix}
    s+1 \\
    m
  \end {pmatrix}
  (k+1-m)^s\ .
\end{equation}
The Eulerian numbers are often denoted as
$\genfrac{\langle}{\rangle}{0pt}{}{s}{k} =A(s,k)$ and they were introduced originally in a combinatorial context.    
Note that $A(s,s)=0$ which is the reason that the sum in \eqref{Eulerpolynomial}
is often extended to include the term $k=s$.
Note that \eqref{Li_recursion} implies that near the pole, as $z \to 1$ one has
$\mathrm{Li}_{-s}(z) \sim \frac{s!}{(1-z)^{s+1}}$ 
which shows that
\begin{equation}
  A_s(1)= \sum_{k=0}^{s-1} A(s,k) =s!\ .
  \label{sum_Eulerian_numbers}
\end{equation}
As the Eulerian numbers are real one also has
$\mathrm{Li}_{-s}(z)^*= \mathrm{Li}_{-s}\left(z^*\right)$.

For the sake of completeness we {  mention} yet another alternative
expression for the trace formula which may be obtained by
expressing $\mathrm{Li}_{-n}(z)$ {in   \eqref{tracepoly1} and
  \eqref{trace_density1}}
as a polynomial in $\frac{z}{1-z}$
\begin{equation}
  \mathrm{Li}_{-n}(z) = \sum_{k=0}^n k! S(n+1,k+1) \left ( \frac{z}{1-z}\right )^{k+1}\ .
\end{equation}
Here $S(n,k)$ are the Stirling numbers of the second kind
\begin{equation}
  S(n,k) =\frac{1}{k!}\sum_{j=0}^{k} (-1)^{k-j}
  \begin{pmatrix} 
    k \\
    j
  \end {pmatrix}
  j^n
\end{equation}
The Stirling numbers satisfy $S(n,n)=S(n,1)=1$.
For fixed k and $n \gg 1 ,\ \ S(n,k) \sim \frac{k^n}{k!}$. 

\subsection{The trace formula as a sum over periodic orbits on an underlying graph}
\label{periodicorbitsI}

To the matrix $H$ we may associate the graph $\mathcal{G}_{I}$ with
$N$ vertices. We enumerate the vertices and associate each to one dimension of
the matrix $H$. The 
adjacency matrix $A_I$ of $\mathcal{G}_{I}$ is given by
\begin{equation}
  A_{I, vw}=
  \begin{cases}
    1 & \text{if $H_{vw}\neq 0$,}\\
    0 & \text{if $H_{vw}=0$.}
  \end{cases}
\end{equation}
So that there is an edge connecting two different vertices $v$ and $w\neq v$
if the
corresponding off-diagonal element of $H_{vw}$ does not vanish, and a loop
at vertex $v$ when the corresponding diagonal element $H_{vv}$ does not vanish.
A periodic orbit $p=\overline{v_1v_2\dots v_{n_p}}$ on the graph $\mathcal{G}_I$
consists of a sequence of vertices $v_k$ ($k=1,\dots,n_p$) such
that, for $k=1,\dots n_p$ the vertex  $v_k$ is connected to $v_{k+1}$ by an
edge (that is $A_{I, v_kv_{k+1}}=1$) and $v_{n_p}$ is also connected to $v_1$
($A_{I, v_{n_p}v_{1}}=1$). Different starting points are considered equivalent
($\overline{v_1v_2\dots v_{n_p}}\equiv \overline{v_2\dots v_{n_p}v_1}$)
and the integer $n_p$ is called the length of the periodic orbit.
If $p=\overline{v_1v_2\dots v_{n_p}}$ we will always set
$v_0\equiv v_{n_p}$ and $v_{n_p+1}=v_1$.
A periodic orbit that is not a repetition of a shorter orbit is called
primitive. Every periodic orbit is a repetition of a primitive orbit
with repetition number $r_p$.
To each periodic orbit $p$ of length $n_p$ we now associate the weight
\begin{equation}
  {  \mathcal{W}_{I, p}}= \prod_{k=1}^{n_p} H_{v_{k+1} v_k}
\end{equation}
such that
\begin{equation}
  \mathrm{tr}(H^s)= \sum_{\mathrm{per.orb.}\ p: n_p=s}
  \frac{s}{r_p} {  \mathcal{W}_{I, p}} , 
\end{equation}
where the sum is over all periodic orbits $p$ of length $n_p=s$. The factor
$\frac{s}{r_p}$ counts the different starting points of a periodic orbit: if
$p$ is a $r_p$-fold repetition of the primitive periodic orbit $\hat{p}$ then
$\frac{s}{r_p}$ is the length of the primitive orbit $\hat{p}$. In that case
one also has ${  \mathcal{W}_{I,p}=\mathcal{W}_{I,\hat{p}}^{r_p}}$.
It is then straight
forward to write the oscillatory parts of the trace formulas
\eqref{tracepoly1_osc} and
\eqref{trace_density1_osc}
as sums over primitive periodic orbits and their repetitions
\begin{align}
  \label{tracepoly2}
  \mathcal{N}_I^{(\mathrm{osc})}(\lambda)=
  &
    \frac{1}{\pi} \lim_{\epsilon \rightarrow  0^+}\Im \left \{
    \sum_{\mathrm{prim.per.orb}\ p} \sum_{r=1}^\infty\
    \frac{n_p\
    e^{(i \lambda -\epsilon)-i\frac{\pi}{2}r n_p}\ A_{r n_p -1}\!\left(e^{(i \lambda
    -\epsilon)}\right)}{
    (rn_p)!\ \left(1-e^{(i \lambda-\epsilon)}\right)^{r n_p}}  
    {  \mathcal{W}_{I, p}^r}
    \right \}
\end{align}
and
\begin{align}
  \label{trace_density2}
  \rho_I^{(\mathrm{osc})}(\lambda)=
  &
    \frac{1}{\pi} \lim_{\epsilon\rightarrow 0^+} \Re
    \left \{
    \sum_{\mathrm{prim.per.orb}\ p} \sum_{r=1}^\infty\
    \frac{n_p\
    e^{(i \lambda -\epsilon)-i\frac{\pi}{2}r n_p}\ A_{r n_p }\!\left(e^{(i \lambda-\epsilon)}\right)}{
    (rn_p)!\ \left(1-e^{(i \lambda-\epsilon)}\right)^{r n_p+1}}
    {  \mathcal{W}_{I, p}^r}
    \right \}\ .
\end{align}
Note that we have expressed \eqref{tracepoly2} and
  \eqref{trace_density2} in terms of Eulerian polynomials rather than the (equivalent) Polylogarithm.

\subsection{Application to spectral averages and the Worpitzky identity}
\label{sec:application_spectral_averages}

Let us now consider how the trace formula in either the form
\eqref{trace_density}
or \eqref{trace_density1} may be used to
perform spectral averages.
For a function $f(\lambda)$ defined on the interval $-\pi < \lambda < \pi$
we define the spectral average as
\begin{equation}
  \left\langle f \right \rangle: = \frac{1}{N} \sum_{j=1}^N f(\lambda_j)
  = \frac{1}{N} \int_{-\pi}^\pi f(\lambda) \rho(\lambda)\
  {  d\lambda}\ .
  \label{spectral_average}
\end{equation}
As we have discussed above the trace formulas
\eqref{trace_density}
and \eqref{trace_density1} are only valid in
a weak sense and thus, in order to use it in  evaluation
of spectral averages,
the function $f(\lambda)$ must belong to the set of test
functions for which
the trace formula is valid. So let us assume that $f(\lambda)=
\sum_{n=-\infty}^\infty f_n e^{in \lambda}$ with coefficients $f_n$ that
decay at least as fast as $e^{-\pi n}$ as $|n| \to \infty$.
If $f(\lambda)$ is a real function then $f_0$ is real and $f_n=f_{-n}^*$
for $n\ge 1$. 
We will consider
$f(\lambda)$ as a function on the unit circle $z=e^{i \lambda}$ by
defining
\begin{equation}
  \hat{f}(z)= \sum_{n=-\infty}^\infty f_n z^n
  \label{analytic_continuation}
\end{equation}
such that $f(\lambda)= \hat{f}(z)$ if $z=e^{i\lambda}$.
Using the trace formula we may now derive the following identity
\begin{equation}
  \langle f\rangle = f(0) +\sum_{s=1}^\infty e^{i s\pi/2}\frac{\tr H^s}{N(s!)^2}
  \frac{d^s}{dz^s}\left[
    A_s(z)\hat{f}(z)
  \right]_{z=1}
  \label{average}
\end{equation}
and prove that the sum over $s$ converges absolutely.
One may derive \eqref{average} 
by replacing $\rho(\lambda)\equiv \rho_I(\lambda)$
by the trace formula
\eqref{trace_density} (or, equivalently \eqref{trace_density1}).
We will consider the individual terms in the sum over $s$
separately and write
\begin{equation}
  \langle f \rangle = F_0 + \lim_{\epsilon\to 0}
  \sum_{s=1}^\infty F_{s,\epsilon}
  \ .
\end{equation}
Here 
\begin{align}
  F_0=
  & \frac{1}{2\pi}\lim_{\epsilon \to 0}\int_{-\pi}^\pi \left(
    1+\sum_{n=1}^\infty e^{- n \epsilon}
    \left(e^{in\lambda} + e^{-in\lambda} \right)
    \right) f(\lambda)\ d\lambda \\
  &= f_0 +\sum_{n=0}^\infty \left(f_n+ f_{-n}\right)= f(0)\ .
\end{align}
The contribution to $\langle f\rangle$ from $s>0$ 
is given by
\begin{align}
  F_{s,\epsilon}=
  & \frac{\tr H^s}{N 2 \pi s!}
    \sum_{n=1}^\infty
     n^s e^{-n \epsilon} \int_{-\pi}^\pi \left(
    e^{-i\pi s/2}e^{i\lambda n}
    +e^{i\pi s/2}e^{-i\lambda n}
    \right)f(\lambda)d\lambda\\
  =
  &
    \frac{\tr H^s}{N s!}
    \sum_{n=1}^\infty
    n^s e^{-n \epsilon} \left(
    e^{-i\pi s/2} f_{-n}
    +e^{i\pi s/2} f_{n}
    \right) \ .
\end{align}
Next let us show
\begin{equation}
  \lim_{\epsilon \to 0} \sum_{s=0}^\infty F_{s,\epsilon}=
  \sum_{s=0}^\infty F_{s,0}
  \label{convergence}
\end{equation}
with absolute convergence of the right-hand side.
For this we consider
\begin{align}
  \sum_{s=0}^\infty \left| F_{s,\epsilon}\right|\le
  C \sum_{s=0}^\infty\sum_{n=1}^\infty
  \frac{n^s \mathrm{max}_{j=1}^N(|\lambda_j|^s)}{s!}e^{-n (\pi+\epsilon)}=
  C \frac{e^{-\epsilon -\delta}}{1- e^{-\epsilon- \delta}}
\end{align}
where $\delta>0$ is the gap defined in \eqref{gap}.
Thus the positivity of the gap ensures that we can take
the limit
$\epsilon \to 0$ in \eqref{convergence} term by term and
absolutely convergence of the sum on the right-hand side.\\
What remains to be shown in order
to derive \eqref{average} is the identity
\begin{equation}
  \sum_{n=1}^\infty n^s 
  \left(
    e^{-i s \pi/2} f_{-n} + e^{is\pi/2} f_n \right)
  = \frac{e^{i \pi s/2 }}{s!}\frac{d^s}{dz^s}\left[
    A_s(z)\hat{f}(z)
  \right]_{z=1}\ .
  \label{resummation}
\end{equation}
This can be shown using the  Worpitzky  identity \cite{worp}
\begin{equation}
  \sum_{k=0}^{s-1} A(s,k) \binom{z+k}{s}=z^s
  \label{worpitzky}
\end{equation}
which is valid for $z \in \CC$ and integer $s\ge 1$.
This identity implies many useful combinatorial relations and is central
to making sense of the trace formulas \eqref{tracepoly1} and \eqref{trace_density1}.
Using \eqref{Eulerpolynomial} and \eqref{analytic_continuation},
and then applying the Worpitzky identity \eqref{worpitzky} one obtains
\begin{align}
  \frac{1}{s!} \frac{d^s}{dz^s}\left[A_s(z)\hat{f}(z)\right]_{z=1}
  =
  &
    \sum_{n=-\infty}^\infty f_n \sum_{k=0}^{s-1} A(s,k) \binom{k+n}{s}
  \nonumber \\
  =
  &
    \sum_{n=-\infty}^\infty f_n n^s=
    \sum_{n=1}^\infty n^s \left(f_n + (-1)^s f_{-n}  \right)
\end{align}
which is equivalent to \eqref{resummation} 
(after multiplication with $e^{i s\pi/2}$). This finishes the derivation
of \eqref{average}. If $f(\lambda)$ is a real test function then the average
$\langle f \rangle$ must be real as well and this is not obvious in the
right-hand side of \eqref{average}. Our derivation shows that the contribution
for each $s$ is individually real. Indeed, for real $f(\lambda)$ one has
$f_n=f_{-n}^*$ and this turns the left-hand side of \eqref{resummation} real.

An alternative derivation of \eqref{average} may be obtained
starting directly
from the resummed trace formula \eqref{trace_density1}.
{  Considering the integral \eqref{spectral_average}
  as an integral over the unit circle $z= e^{i \lambda}$
  in the complex plane, the} Eulerian
polynomials are directly related to the expansion of the Polylogarithm
around their pole at $z=1$.

While \eqref{average} is valid rigorously only for
very restricted test functions let us now illustrate that
it can be applied formally to a larger set of test functions
and give formally correct results.
For this we consider the heat kernel
\begin{equation}
  \frac{1}{N}\tr [e^{-\beta H}]= \langle e^{-\beta \lambda} \rangle \ .
  \label{heat}
\end{equation}
This identity is obvious when
we express $\rho(\lambda)=\sum_{j=1}^N \delta(\lambda-\lambda_j)$
in
$\langle e^{-\beta \lambda} \rangle= \frac{1}{N} \int_{-\pi}^\pi e^{-\beta
  \lambda} \rho(\lambda) d\lambda$.  
We will show that the trace formula gives back that result as well.
So let $f(\lambda)=e^{-\beta \lambda}$ for $\lambda \in (-\pi,\pi)$.
It has Fourier coefficients $f_n$ which only decay as $1/|n|$. This is too
weak to ensure convergence of the trace formula and
$f(\lambda)$ is formally not an allowed test function.
Inserting $f(\lambda)=e^{-\beta \lambda}$ (and thus $\hat{f}(z)= z^{i \beta}$)
into \eqref{average} one may use the Worpitzky identity \eqref{worpitzky}
again to evaluate
\begin{equation}
  \frac{d^s}{dz^s}\left[ A_s(z) z^{i \beta}\right]_{z=1}=
  \sum_{m=0}^{s-1} A(s,m)\prod_{k=1}^s((m+1-k)+i\beta)\
  = s! \sum_{m=0}^{s-1} \binom{m+i\beta}{s}
  s! (i\beta)^s \ .
\label{identity}
\end{equation}
We then find
\begin{align}
  \langle f \rangle =
  1 + \frac{1}{N} \sum_{s=1}^\infty \frac{(-\beta)^s \mathrm{tr}\ H^s}{s!}
  =
  \frac{1}{N}\tr [e^{-\beta H}]
\end{align}
as expected.

Another instructive example is to consider
$f(\lambda) = \lambda^m$ for integer $m$ and $\lambda \in (-\pi,\pi)$.
Again the Fourier coefficients only decay as $1/|n|$. Nonetheless
we will show that \eqref{average} formally
recovers $\langle f\rangle = \frac{1}{N} \mathrm{tr}\ H^m$.
Substituting  $\hat{f}(z)=(-i\log z)^m$
into the right-hand side we thus need to show
\begin{eqnarray}
  \left \{ e^{i\frac{\pi}{2}(s-m)}
  \frac{1}{(s !)^2}\ \left [
  \frac{d^s\  }{dz^s}\left ( A_s(z)(\log z)^m\right )
  \right ]_{z=1}
  \right \}\ = \ \delta_{s,m}\ .
  \label{meanfy1}
\end{eqnarray}
This is equivalent to requiring
\begin{equation}
  \frac{1}{(s !)^2}\sum_{k=0}^{s-1} A(s,k)
  \left [\frac{d^s}{d z^s}\left( z^k(\log z)^m\right )  \right ]_{z=1}
  =  \delta_{s,m}
  \label{identity3}
\end{equation}
This identity is derived in Appendix~\ref{appendix} alongside a number
of other
useful relations that follow from the Worpitzky identity.

\subsection {Example: 
  Derivation of the semi-circle law from Wigner's estimate of
  $\left \langle  \tr H ^n\right \rangle$
}

Another possible application of the trace formulas \eqref{tracepoly},
\eqref{trace_density}, \eqref{tracepoly1} and \eqref{trace_density1}
that we will now explore is in random-matrix theory.
For instance one choose $H$ to be a random element of one of
the Gaussian $\beta$ ensembles G$\beta$E (for $\beta=1$ and
$\beta=2$ this is GOE and GUE) \cite{qsoc,mehta,forrester,betaensembles}.
We will denote the ensemble average of any function $f(H)$ as
$\langle f(H)\rangle_\beta$.

Note that these matrices generally do not obey the
restriction that the
spectrum is contained in $(-\pi,\pi)$.
So in general $\mathcal{N}(\lambda) \neq \mathcal{N}_I(\lambda)$ even for $|\lambda|<\pi$.
As is well known,
in the limit $N \to \infty$ of
the ensemble average of the density of states 
is given by Wigner's semicircle law which
by appropriate scaling
(which coincides with a standard convention
in random-matrix theory) 
limits the spectrum of the ensemble element $H$
to the interval $[-\pi,\pi]$ with unit
probability. For finite values of $N$ the average
density of states has
tails which extend to arbitrarily large values
of $|\lambda|$.
We will show that the trace formulas built on the
evolution operator
$S_I(\lambda)= e^{i(\lambda-H)}$ give a consistent
relation
between Wigner semi-circle law and the
known asymptotic formulas
for ensembles averaged traces
$\left \langle \mathrm{tr}\ H^s\right \rangle_\beta$.
%By direct expansion of the Wigner semi-circle law
%into traces of $H$ the large $N$ asymptotics of the
%latter is given by
{ 
  The following derivation of the Wigner semi-circle law for the
  $\beta$-ensembles is analogous to
  Wigner's derivation \cite{wigner_bernouilli} of the same law for random Bernouilli matrices
  (symmetric matrices with matrix elements equal to zero or one with
  probability one half). 
 The large $N$ expansion of these traces may be obtained from
 recursion formulas \cite{GUE_traces} and is given by
}
\begin{equation}
  \frac{1}{N}\left \langle \tr H^{2p}\right \rangle_{\beta}\sim \frac{(2p)!}{p!(p+1)!}\left (\frac{\pi}{2}\right )^{2p}
  \label{trace_asymptotics}
\end{equation}
for even powers while for odd powers the trace vanishes
trivially.
{  Substituting this into the averaged trace
formula  \eqref{tracepoly} for $\mathcal{N}_I(\lambda)$ we 
will now rederive 
the integrated Wigner semi-circle law by resummation.}
%We will show that this is consistent with the trace
%formula  \eqref{tracepoly} for $\mathcal{N}_I(\lambda)$
%by starting from
%\eqref{trace_asymptotics} and substituting this into
%$\mathcal{N}_I(\lambda)$
%and resumming the (integrated version of)
%Wigner semi-circle law using the trace formula.
Using
\begin{equation}
  \sum_{p=0}^{\infty} (-1)^p\frac{1}{p!(p+1)!}\left (\frac{\pi}{2}\right )^{2p}n^{2p-1}=\frac{2}{\pi n^2}J_1(n \pi),
\end{equation}
one obtains
\begin{align}
  \frac{1}{N}
  \langle  \mathcal{N}_I(\lambda) \rangle _{\beta} \sim
  &
    \frac{1}{2}+ \frac{\lambda}{2\pi}
    +\frac{2}{\pi^2}\sum_{n=1}^{\infty} J_1(n\pi) \frac{\sin (n\lambda)}{n^2}\ . 
    \label{scirclen}
\end{align}
Note that the sum above is just the Fourier transform of the
oscillating part. Using \eqref{tracepoly} the latter may also be written
as
\begin{equation}
  \left\langle \mathcal{N}_I^{(\mathrm{osc})}(\lambda) \right\rangle=
  \sum_{n=1}^\infty\frac{1}{n \pi}\left\langle \tr
    \left[\cos(n H)\right] \right\rangle_\beta
  \sin(n \lambda)
\end{equation}
which shows that the trace formula \eqref{tracepoly1} for
$\mathcal{N}_I(\lambda)$ implies
\begin{equation}
  \frac{1}{N}\left\langle \tr [\cos(n H)] \right\rangle_{\beta} =
  \frac{1}{N}\left\langle \tr [e^{inH}] \right\rangle_{\beta}
  \sim \frac{2}{n\pi}J_{1}(n\pi)\ .
\end{equation}
In order to show that these results are 
%consistent with 
{  equivalent to}
the integrated  semi-circle distribution
\begin{equation}
  \frac{1}{N}\left\langle \mathcal{N}(\lambda)
  \right\rangle_{\beta, \textrm{semi-circle}}=
  \frac{1}{2} +\frac{1}{\pi}\left [ \frac{\lambda}{\pi}  \sqrt{1-
      \left (\frac{\lambda}{\pi}\right )^2}
    + \arcsin\left(\frac{\lambda}{\pi}\right)\right ]
  \label{scircle}
\end{equation}
one needs to compute the $\sin$-transform of
$\frac{1}{N}\left\langle \mathcal{N}(\lambda) \right\rangle_{\beta, \mathrm{semi-circle}}-\frac{\pi+\lambda}{2\pi}$
and show that it is consistent with the result obtained from
the trace formula \eqref{scirclen}.
This follows by using Hankel's  integral expression
\begin{equation}
  J_1(n\pi)=n \int_{-1}^1  \sqrt{ 1-t^2}  \cos (n \pi t)\ {  d}t \ .
\end{equation}

\section{Approach II: trace formula from an evolution operator
  on the directed
  edges of the associated simple graph}
\label{tf2}

In this section let $H$ be an arbitrary $N \times N$ Hermitian matrix.
We do not require the the spectrum of $H$ is restricted to any interval.
In Sec.~\ref{periodicorbitsI} we have written the trace formulas
\eqref{tracepoly2} and \eqref{trace_density2} as sums over contributions from
periodic orbits on the underlying graph $\mathcal{G}_I$.
This graph
generally contains
loops which stand for the (non-vanishing) diagonal elements of the matrix $H$
and it is thus generally not simple.
In this section we 
use a different approach that starts from an associated
simple graph $\mathcal{G}_{II}$ that is obtained from $\mathcal{G}_I$ by
taking away the loops.
So $\mathcal{G}_{II}$ has again $N$ vertices
and two \emph{different} vertices $v\neq w$ are connected
%and form an edge $(v,w)$
%of $\mathcal{G}_H$
if the correspondent non-diagonal matrix element
$H_{vw}$ does not vanish. The adjacency matrix of $\mathcal{G}_{II}$ is then
\begin{equation}
  A_{vw}=
  \begin{cases}
    1& \text{if $v\neq w$ and $H_{vw}\neq 0$}\\
    0& \text{if $v=w$ or $H_{vw}=0$.}
  \end{cases}
\end{equation}
If $H$ is a full matrix (or has vanishing entries only on the diagonal)
the associated graph is the complete graph on $N$ vertices.
The number of edges of the associated graph is given by
$E=\frac{1}{2}\sum_{v,w=1}^N A_{vw}$.
For later use let us introduce the neighbourhood
{  $ \mathcal{E}_v$ } %$\mathcal{N}_v$
of a vertex $v$ in the graph as the set of vertices $w\neq v$ 
that are connected to $v$ by an edge. The degree $d_v$
(also known as valency or coordination number) of the vertex $v$
is the number of adjacent edges or, equivalently the number or neighbour
vertices
$d_v={  \left|\mathcal{E}_v\right|}$.

When $A_{vw}=1$ we write
\begin{equation}
	H_{vw}= H_{wv}^*= h_{vw}e^{2i\gamma_{vw}}
\end{equation}
where $v,w \in \{1,2,\dots,N\}$, $h_{vw}=h_{wv}=\mathrm{abs}(H_{vw})\ge 0$
and $\gamma_{vw}=-\gamma_{wv}\in [-\pi/2, \pi/2]$.
If $H_{vw}$ is real and negative we choose $\gamma_{vw}= \pi/2$ if
$v \ge w$ and $\gamma_{vw}=-\pi/2$ if $v < w$.\\
Our aim will be to rewrite the corresponding eigenproblem 
\begin{equation}
  \sum_{w=1}^N H_{vw} \phi_w= \lambda \phi_v
  \qquad \text{for $v\in\{1,2,\dots,N\}$}
\end{equation}
with a real spectral parameter $\lambda$ as a unitary
scattering problem on
the associated graph.
This will allow us to derive a trace formula that expresses
the spectrum of the matrix in terms of periodic orbits on the associated
graph $\mathcal{G}_{II}$.
This trace formula will turn out to be different from
the one
derived in Section~\ref{tf1}.
The first difference is that the new trace formula
will describe $\mathcal{N}(\lambda)$ for $\lambda$ on the
real line.
Another difference is in the set of periodic orbits   
which is larger in the first approach by containing
additional loops
(repeated vertices). And one more difference is that the
weight of a periodic orbit
in Section~\ref{periodicorbitsI} contains the product
of matrix
elements of $H$ along the
orbit while here we will derive a construction
where the weight is a product
of scattering amplitudes that stem from
unitary scattering matrices at
the vertices along the orbit.

\subsection{Wave function amplitudes on directed edges}

For any edge $(v,w)$ we introduce two
complex amplitudes $a_{vw}$ and $a_{wv}$. 
One may think of
$a_{vw}$ as the amplitude of a wave going from vertex $w$ to vertex $v$,
and of $a_{wv}$ as an amplitude for a counter-propagating wave from $v$ to $w$.
This physical interpretation is often helpful but
not necessary for 
{  the following construction.}
We will however use 
the double indices $vw$ and $wv$
{  to distinguish between the}
two
directions on the edge $(v,w)$. 
Next we
{  express the $N$ complex vector components $\phi_v$ as}
a linear combination of {  complex amplitudes $a_{vw}$ and $a_{wv}$ on
  adjacent edges (that is $w \in \mathcal{E}_v$) } 
\begin{equation}
  \phi_v=
  \frac{e^{i \gamma_{vw}}}{\sqrt{h_{vw}}}
  \left[a_{vw}e^{- i \pi/4}+ a_{wv} e^{i \pi/4} \right]\ .
  \label{from_vertex_to_edge_amplitudes}
\end{equation}
On a given edge $(v,w)$ the definition \eqref{from_vertex_to_edge_amplitudes}
implies (by swapping indices $v$ and $w$)
\begin{equation}
  \begin{split}
    \phi_w=&\frac{e^{i \gamma_{wv}}}{\sqrt{h_{wv}}}
    \left[a_{wv}e^{-i \pi/4 }+ a_{vw} e^{i\pi/4} \right]\\
    =& \frac{e^{-i \gamma_{vw}}}{\sqrt{h_{wv}}}
    \left[a_{vw} e^{i\pi/4}+a_{wv}e^{- i \pi/4}\right] .
  \end{split}
  \label{from_other_vertex_to_edge_amplitudes}
\end{equation}
The physical interpretation of \eqref{from_vertex_to_edge_amplitudes}
and \eqref{from_other_vertex_to_edge_amplitudes} is that the wave with
amplitude $a_{vw}$ travels from vertex $w$ to $v$ and acquires an additional
phase $-ie^{2i \gamma_{vw}}$
while the counter-propagating wave
acquires the phase
$-i e^{2i\gamma_{wv}}$.
The two phases are different for $\gamma_{vw}\neq 0{ , \pi/2}$ as
$e^{2i\gamma_{wv}}=e^{-2i \gamma_{vw}}\neq e^{2i\gamma_{vw}}$.

\subsection{Vertex scattering matrices}
\label{sec:vertex_scattering}

At a given vertex $v$ with valency $d_v$ we
have $d_v$ incoming wave amplitudes $a_{vw}$ and $d_v$
outgoing amplitudes $a_{wv}$.
Our next aim is to
derive a linear
{  relation between} the outgoing amplitudes on the incoming amplitudes
that contains all relevant information on the spectrum of the matrix $H$.
We thus need $d_v$ linear relations between the wave amplitudes. To proceed,
first note that we can write
$\phi_v$ in $d_v$ different ways using
\eqref{from_vertex_to_edge_amplitudes} (one for each edge adjacent to
$v$). This gives $d_v-1$ independent linear relations
\begin{equation}
  \begin{split}
    \phi_v=&
    \frac{ e^{ i\gamma_{vw} } }{\sqrt{ h_{vw} }}
    \left[
      e^{-i\pi/4} a_{vw}+e^{i\pi/4}a_{wv}	
    \right]
    \\
    =& 
    \frac{ e^{ i\gamma_{vw'} } }{ \sqrt{ h_{vw'} }}
    \left[
      e^{-i\pi/4} a_{vw'}+e^{i\pi/4}a_{w'v}	
    \right]
    \\
    =& 
    \frac{1}{d_v}\sum_{w\in {  \mathcal{E}_v}}    \frac{e^{ i\gamma_{vw} }}{\sqrt{h_{vw}}}
    \left[
      e^{-i\pi/4} a_{vw}+e^{i\pi/4}a_{wv}	
    \right]
  \end{split}
  \label{eq:continuity}
\end{equation}
We get one more condition by taking the $v$-th equation of the eigenproblem
\begin{equation}
  (H_{vv}-\lambda)\phi_v + \sum_{w\in {  \mathcal{E}_v}} H_{vw} \phi_w=0
\end{equation}
and expressing the vertex amplitudes by linear combinations of wave
amplitudes on the adjacent edges using \eqref{from_vertex_to_edge_amplitudes}
and
\eqref{eq:continuity}.
This results in the equation
\begin{equation}
  \frac{H_{vv}-\lambda}{d_v}
  \sum_{w\in {  \mathcal{E}_v}}
  \frac{e^{ i\gamma_{vw} }}{\sqrt{h_{vw}}}
  \left[
    e^{-i\pi/4} a_{vw}+e^{i\pi/4}a_{wv}	
  \right]
  =
  - 
  \sum_{w\in {  \mathcal{E}_v}} 
  H_{vw} 
  \frac{e^{ i\gamma_{wv} }}{\sqrt{h_{wv}}}
  \left[
    e^{-i\pi/4} a_{wv}+e^{i\pi/4}a_{vw}	
  \right]
  \label{eq:eigen}
\end{equation}
where the sums over $w$ extend over the $d_v$ vertices $w$ adjacent to $v$.
We may solve \eqref{eq:continuity} and \eqref{eq:eigen} 
and express the outgoing amplitude $a_{wv}$ as
\begin{equation}
  \begin{split}
    a_{wv }=& i a_{v w}
    -2\sum_{w'\in {  \mathcal{E}_v}} \frac{\sqrt{h_{vw'}h_{vw}}}{
      H_{vv}-\lambda-i\Gamma_v} 
    e^{i(\gamma_{vw'}+\gamma_{wv})} a_{vw'}\\
    =&
    i\frac{H_{vv}-\lambda-i\Gamma_v
      +2i h_{vw}}{H_{vv}-\lambda -i\Gamma_v}a_{v w}\\
    &
    -2 \sum_{w'\in {  \mathcal{E}_v}, w' \neq w} \frac{\sqrt{h_{vw'}h_{vw}}}{
      H_{vv}-\lambda-i\Gamma_v} 
    e^{i(\gamma_{vw'}+\gamma_{wv})} a_{vw'}
  \end{split}
  \label{eq:outgoing}
\end{equation}
where
\begin{equation}
  \Gamma_v= \sum_{w \in  {  \mathcal{E}_v}   } h_{vw} {  \ .}
\end{equation}

Writing \eqref{eq:outgoing} as
\begin{equation}
  a_{wv} = \sum_{w' \in  {  \mathcal{E}_v}} \sigma_{w w'}^{(v)}(\lambda)a_{v,w'}
\end{equation}
one obtains a $d_v \times d_v$ vertex scattering matrix
$\sigma^{(v)}(\lambda)$. One may write this as
\begin{equation}
  \sigma^{(v)}(\lambda)=i \II-\frac{2}{H_{vv}-\lambda-i\Gamma_v}
  \boldsymbol{\Lambda}^{(v)} \boldsymbol{\Lambda}^{(v) \dagger}
  \label{eq:vertex_scattering_matrix}
\end{equation}
where
$\boldsymbol{\Lambda}^{(v)}$ is a $d_v$-dimensional column vector with elements
\begin{equation}
  \Lambda^{(v)}_{w}= \sqrt{h_{vw}}e^{-i \gamma_{vw}}\ .
  \label{couplingconstants}
\end{equation}
Using 
$\boldsymbol{\Lambda}^{(v)\ \dagger} \boldsymbol{\Lambda}^{(v)}=\Gamma_v
{  \II}$ 
one may show unitarity of the vertex scattering matrix
\begin{equation}
  \sigma^{(v)}(\lambda)^\dagger \sigma^{(v)}(\lambda)=
  \sigma^{(v)}(\lambda)\sigma^{(v)}(\lambda)^\dagger= \II
\end{equation}
with a straight forward calculation.

{  The expression \eqref{eq:vertex_scattering_matrix}
  brings together two concepts. Note first that $\Gamma_v$
  is known as the Gershgorin radius of Gershgorin's circle theorem
  \cite{gershgorin} which (for the present
  context) states that each eigenvalue $\lambda$ of the Hermitian matrix
  ${ H}$ lies
  in at least one of the $n$ intervals (`discs')
  $[H_{vv}-\Gamma_v,H_{vv}+\Gamma_v]$ ($v=1,\dots, N$).\footnote{Gershgorin's
    theorem \cite{gershgorin}
    applies more generally to complex matrices $A$ where it states that
    each
    (generally complex) eigenvalue $\lambda$ of $A$ lies in at least one of the
    Gershgorin discs $|A_{ii}-\lambda|\le \Gamma_i= \sum_{j\neq i} |A_{ij}|$.
    Note that it is
    in general not true that each  Gershgorin disc contains at least one
    eigenvalue but if a Gershgorin disc is disjoint from the union of all other
    discs then it must contain one eigenvalue.
  }
  Second, the vertex scattering matrix
  \eqref{eq:vertex_scattering_matrix} is of the
  standard form derived by Weidenm\"uller and others (see e.g.
  \cite{weidenmuller}) for
  scattering from a system with a single bound state (with unperturbed energy
  $H_{vv}$ coupled to $d_v$ channels with coupling constants \eqref{couplingconstants}.)
}

The expression \eqref{eq:vertex_scattering_matrix} also allows for a
calculation of the complete
spectrum of $\sigma^{(v)}(\lambda)$.
Obviously,  $\boldsymbol{\Lambda}^{(v)}$
is an eigenvector of the vertex scattering matrix
\begin{equation}
	\sigma^{(v)}(\lambda)\boldsymbol{\Lambda}^{(v)}=
        i \frac{H_{vv}-\lambda+i\Gamma_v}{H_{vv}-\lambda-i\Gamma_v}
        \boldsymbol{\Lambda}^{(v)}\ .
\end{equation}
Next, let $\mathbf{u}$ be any vector orthogonal to $\boldsymbol{\Lambda}^{(v)}$
, i.e.
$\mathbf{u}^\dagger \boldsymbol{\Lambda}^{(v)}=0$. 
Then
\begin{equation}
	\sigma^{(v)}(\lambda) \mathbf{u} = i \mathbf{u}
\end{equation}
with eigenvalue $i$. The eigenvalue $i$ is $d_v-1$-fold
degenerate as there are $d_v-1$
linearly independent choices of $\mathbf{u}$ orthogonal to
$\boldsymbol{\Lambda}^{(v)}$.
It follows that the determinant of the vertex scattering matrix is given by
\begin{equation}
  \det\ \sigma^{(v)}(\lambda) = i^{d_v}\frac{H_{vv}-\lambda+i\Gamma_v}{H_{vv}-\lambda-i\Gamma_v}
\end{equation}
which can also be calculated directly from
\eqref{eq:vertex_scattering_matrix}.

{  Note, that if one matrix element $H_{v\hat{w}}$ is very small
then also the corresponding component $\Lambda^{(v)}_{\hat{w}}$ becomes small
which suppresses the modulus of 
scattering amplitude between the edge $(v,\hat{w})$ and any other edge
$v$ while backscattering is increased.
In the limit $H_{v\hat{w}} \to 0$ one finds indeed (for $w \neq\hat{w}$)
$\sigma^{(v)}_{\hat{w} \hat{w}}(\lambda) \to i$,
$\sigma^{(v)}_{\hat{w}w} \to 0$ and $\sigma^{(v)}_{w\hat{w}} \to 0$ which
effectively decouples the edge $(v,\hat{w})$ from the vertex $v$.
At the vertex $\hat{w}$ the edge $(v,\hat{w})$ decouples in the same
way and the corresponding coefficients have to vanish
$a_{v\hat{w}}=0= a_{\hat{w}v}$ making this edge redundant.}

{  Some simple cases are computed explicitly in App.~\ref{appendix_examples} to
  illustrate the discussion above.}

\subsection{The discrete-time quantum evolution operator 
 and its spectral determinant}

Writing all $2E$ coefficients $a_{vw}$ as a column vector $\mathbf{a}$
one may now write all the
matching conditions at all vertices as
\begin{equation}
  \mathbf{a}= S_{II}(\lambda) \mathbf{a}
  \label{eq:scattering}
\end{equation}
using a single $2E \times 2E$ dimensional unitary matrix
\begin{equation}
  S_{II\ v'w',vw}(\lambda)= \delta_{w' v} \sigma^{(v)}_{v'w}(\lambda)
  \label{eq:quantum_map}
\end{equation}
where the double indices $vw$ or $v'w'$ run over the $2E$ directed edges.
We will call $S_{II}(\lambda)$ the discrete-time quantum
evolution operator.
For clarity we would like to stress that the discrete time steps do not
correspond to a discretization of a continuous time. It is a
``topological'' time that counts the number of scattering events.
One may view  $S_{II}(\lambda)$ as the evolution operator
of a discrete-time quantum walk on the directed edges
{  (see Sect.~\ref{quantum_walks})}
and its unitarity can be
verified straight-forwardly by observing that
(with an appropriate choice of order of the amplitudes $a_{vw}$ in the vector $\mathbf{a}$)
one may write
\begin{equation}
  S_{II}(\lambda)= {  P }
  \begin{pmatrix}
    \sigma^{(1)}(\lambda) & 0 & \dots & 0\\
    0 & \sigma^{(2)}(\lambda) & \dots & 0\\
    \dots&\dots&\dots& \dots \\
    0 & 0 & \dots & \sigma^{(N)}(\lambda)
  \end{pmatrix} \ .
\end{equation}
{  Here 
  $P$ is the permutation matrix} that interchanges each directed
edge with the opposite direction on the same
edge, and {  thus $P^2=\mathbb{I}$}.
The unitarity of $S_{II}(\lambda)$ then follows from the unitarity of
{  the permutation matrix $P$}
and the vertex scattering matrices $\sigma^{(v)}(\lambda)$
that 
appear as diagonal blocks. The determinant of $S_{II}(\lambda)$
may be calculated straight-forwardly. 
As {  $P$} consists of
$E$ transpositions one obtains
\begin{equation}
  \det S_{II}(\lambda) =
	(-1)^E \prod_{v=1}^N \det\ \sigma^{(v)}(\lambda)= \prod_{v=1}^N 
	\frac{H_{vv}-\lambda+i \Gamma_v}{H_{vv}-\lambda-i\Gamma_v} \
        {  .}
\end{equation}

The set of linear equations \eqref{eq:scattering} has non-trivial solutions if
the corresponding determinant $\det \left( \II - S_{II}(\lambda) \right)=0$.
Let us thus define the spectral determinant
\begin{equation}
  \zeta_{II}(\lambda)=\det \left( \II - S_{II}(\lambda) \right)
  \label{eq:spectraldeterminant_II}
\end{equation}
such that there is a non-trivial solution to \eqref{eq:scattering} if
\begin{equation}
  \zeta_{II}(\lambda)=0\ .
\end{equation}
Let us denote the set of real 
values of the spectral parameter where this occurs as
\begin{equation}
  \varsigma_{II}(H)= \left\{ \lambda \in \mathbb{R}: \zeta_{II}(\lambda)=0
  \right\} .
\end{equation}
Denoting the spectrum of $H$ as $\varsigma(H)$
one has $\varsigma(H) \subset \varsigma_{II}(H)$ because the conditions
\eqref{eq:scattering} are satisfied by construction if
$\lambda \in \varsigma(H)$.
The identity
$\varsigma(H) = \varsigma_{II}(H)$ follows from the { 
  determinantal}
identity
\begin{equation}
  \zeta_{II}(\lambda)=
  \frac{2^E \zeta_H(\lambda)}{\prod_{v=1}^N (H_{vv}-\lambda - i \Gamma_v)}
  \label{eq:factorisation}
\end{equation}
which we are now going to derive.
For this we consider $\zeta_{II}(\lambda)$ as a complex function with
$\lambda \in \CC$. By construction $\zeta_{II}(\lambda)$ is a rational
function of $\lambda$ and
$\zeta_H(\lambda)=0$ implies $\zeta_{II}(\lambda){  =0}$.
{  It follows that one can write}
\begin{equation}
  \zeta_{II}(\lambda)= C \frac{p(\lambda)}{q(\lambda)} \zeta_H(\lambda)
\end{equation}
where $C$ is a constant and $p(\lambda)= \prod_{k=1}^{n_p}(p_k -\lambda)$
and $q(\lambda)= \prod_{l=1}^{n_q}(q_l-\lambda)$ are 
polynomials of degrees
$n_p$ and $n_q$ which we have written in factorized form.
Without loss of generality we
may assume that $p_k \neq q_l$ (for $1\le k \le n_p$ and $1\le l \le n_q$).
By considering $\lambda \to \infty$ one may find a relation between
the orders of the polynomials and fix the constant $C$.
In this limit each vertex scattering matrix becomes proportional to the
identity
$\sigma^{(v)} \to i $ and thus $S_{II}(\lambda) \to i {  P}$.
From this one finds by direct calculation (using that ${  P}$ is the permutation
matrix that interchanges a directed edge with the opposite directed
edge between
the same vertices)
\begin{equation}
  \begin{split}
    \lim_{\lambda \to \infty}
    \zeta_{II}(\lambda) = & \det(\II - i {  P})\\
    =&
    \left(\det
      \begin{pmatrix}
        1& -i\\
        -i & 1
      \end{pmatrix}
    \right)^E
    =
    2^E\ .
  \end{split}
\end{equation}
So, $C= 2^E$ and the orders of the polynomials obey $n_q=n_p+N$.
It remains to be shown that
$q(\lambda)=\prod_{v=1}^N(H_{vv}-i \Gamma_v -\lambda)$
as this implies $n_q=N$, $n_p=0$ and $p(\lambda)=1$. 
It is obvious from the construction that the matrix $S_{II}(\lambda)$
has poles at the positions $\lambda=H_{vv}-i \Gamma_v$ and no other poles.
It is less obvious that these poles are all simple poles when the determinant
is calculated (naively one may be tempted to believe that they come
with a multiplicity $d_v$).
In Sec.~\ref{sec:vertex_scattering} we have calculated the spectrum of the
vertex scattering matrices that enter the evolution matrix
which consists of a $d_v-1$-fold degenerate eigenvalue $i$ and one
non-degenerate eigenvalue
$i \frac{H_{vv}+i \Gamma_v -\lambda}{H_{vv}-i \Gamma_v -\lambda}$ while
the corresponding eigenvectors may be chosen independent of $\lambda$.
Hence we may diagonalize
\begin{equation}
  \sigma^{(v)}(\lambda) = U^{(v)}
  \tilde{\sigma}^{(v)}(\lambda)
  {U^{(v)}}^\dagger
\end{equation}
with
\begin{equation}
  \tilde{\sigma}^{(v)}(\lambda)=
  \begin{pmatrix}
    i \frac{H_{vv}+i \Gamma_v -\lambda}{H_{vv}-i \Gamma_v -\lambda} & 0 & 0 &\dots\\
    0 & i & 0 & \dots\\
    0 & 0 & i& \dots\\
    \dots&\dots&\dots&\dots
  \end{pmatrix}
\end{equation}
where $U^{(v)}$ is a $d_v$-dimensional unitary matrix that does not depend
on $\lambda$. This implies
\begin{equation}
  \zeta_{II}(\lambda)=
  \det\left(\mathbb{I}-  U^{\dagger}PU \tilde{\sigma}(\lambda)  \right)=
  \det(P) \det\left(U^{\dagger}PU - \tilde{\sigma}(\lambda)  \right)
\end{equation}
where $U$ and $\tilde{\sigma}(\lambda)$ are the block-diagonal matrices with
$N$ diagonal blocks $U^{(v)}$ and $\tilde{\sigma}^{(v)}(\lambda)$.
In the matrix $U^{\dagger}PU - \tilde{\sigma}(\lambda)$
the poles at $\lambda= H_{vv}-i \Gamma_v$ appear in a single matrix element
and thus the determinant has single poles as well.
This concludes the derivation of equation \eqref{eq:factorisation}.

It is useful and interesting in its own right to consider the
spectrum of the quantum evolution operator
$S_{II}(\lambda)$ for a given value of
$\lambda$. As $S_{II}(\lambda)$ is a unitary matrix
there are $2E$ eigenvalues of the form $e^{-i\theta_n(\lambda)}$
($n=1,\dots,2E$). 
One may generalize the definition of the spectral determinant
as
\begin{equation}
  \zeta_{II}(\lambda,z) = \det \left( \II - z S_{II}(\lambda) \right)
\end{equation}
such that $\zeta_{II}(\lambda, e^{i\theta})=0$ if and only if
$\theta=\theta_n(\lambda)\ \mathrm{mod}\ 2\pi$ is an eigenvalue
of $S_{II}(\lambda)$.
We will continue to use $\zeta_{II}(\lambda)\equiv \zeta_{II}(\lambda,z=1)$.
{  A few simple examples are given in App.~\ref{appendix_examples}.}

\subsection{The trace formula for the counting function and
  the density of states}

We have shown that the quantum evolution matrix
$S_{II}(\lambda)$ is unitary for real spectral
parameter $\lambda$. If one chooses $\lambda \in \mathbb{C}$ in the complex
plane then $S_{II}(\lambda)$ is in general not unitary.
For sufficiently
small $\epsilon$ one then finds (by direct calculation)
$|\det S_{II}(\lambda+i \epsilon)|<1$
which indicates that $S_{II}(\lambda+ i \epsilon)$
is subunitary.
{  The determinantal form of the secular equation
  \eqref{eq:spectraldeterminant_II} and the
  unitary of $S_{II}(\lambda)$ (for real $\lambda$) allows the use of Cauchy's
  theorem to compute the number counting function in the same manner as it was
  used in the previous chapter.}
Under these conditions one finds that the
spectral counting function is given by
\begin{subequations}
  \label{eq:traceformula_II}
  \begin{align}
    \mathcal{N}_{II}(\lambda) =
    &
      \overline{\mathcal{N}}_{II}(\lambda)+
      \mathcal{N}^{\mathrm{(osc)}}(\lambda) \\
    \overline{\mathcal{N}}_{II}(\lambda)
    =
    &
      \frac{1}{2\pi} 
      \mathrm{Im} \log \det S_{II}(\lambda)
      \nonumber\\
    =&
       \sum_{v=1}^N \frac{1}{\pi} \mathrm{arccos}\left(\frac{H_{vv}-\lambda}{
       \sqrt{(H_{vv}-\lambda)^2+\Gamma_v^2}}\right)
    \\
    \mathcal{N}_{II}^{\mathrm{(osc)}}(\lambda)
    =
    &
      - \frac{1}{\pi} \lim_{\epsilon\to 0^+}
      \mathrm{Im} \log \zeta_{II}(\lambda+i \epsilon)
      \nonumber\\
    =&
       \lim_{\epsilon\to 0^+}  
       \frac{1}{\pi} \mathrm{Im} \sum_{n=1}^\infty \frac{1}{n}
       \mathrm{tr}\ S_{II}(\lambda+i \epsilon)^n    
  \end{align}
\end{subequations}
{  A corresponding trace formula
  $\rho_{II}(\lambda)= \overline{\rho}_{II}(\lambda)+
  \rho_{II}^{\mathrm{(osc)}}(\lambda)$ for the density of states
    is obtained by differentiation, $\overline{\rho}_{II(\lambda)}=
    \frac{d}{d\lambda}\overline{\mathcal{N}}_{II}(\lambda) $
    and
    $\rho_{II(\lambda)}^{\mathrm{(osc)}}=
    \frac{d}{d\lambda}\mathcal{N}_{II}^{\mathrm{(osc)}}(\lambda) $.
  }
Note that 
$\lim_{\lambda \to -\infty}
\overline{\mathcal{N}}_{II}(\lambda) =0$
and
$\lim_{\lambda \to -\infty}
\mathcal{N}_{II}^{\mathrm{(osc)}}(\lambda)=0$.
Here $\overline{\mathcal{N}}_{II}(\lambda)$
describes a smooth increase of the spectral counting
function.
Note that this increase is given in terms
of $N$ separate {  smoothed out} %smeared out
steps centered at the diagonal
matrix elements $H_{vv}$ and a width of the
% smeared
{  smoothed}
step given by the Gershgorin radius $\Gamma_v$. 
In other words the smooth part of this trace formula
contains already some information about the location of the
spectrum and this information is consistent with
Gershgorin's theorem (which bounds the spectrum to intervals
of size $2\Gamma_v$ centered at $H_{vv}$).
This is in contrast to the trace formula for the counting
function $\mathcal{N}_I(\lambda)$
in the first approach where the smooth part
$\overline{\mathcal{N}}_I(\lambda)$ just gives a uniform
linear increase with no information on the location
of the spectrum.

The 
oscillating part is a sum of traces and may thus be recast
as a sum over primitive periodic orbits
and their repetition
\begin{equation}
  \begin{split}
    \mathcal{N}_{II}^{(\mathrm{osc})}(\lambda)
    =&
    \mathrm{Im} \sum_{p} \sum_{r=1}^\infty
    \frac{{  \mathcal{W}_{II,p}^r}}{\pi r} \ .
  \end{split}
  \label{countingfunction_periodicorbitsum}
\end{equation}
In the last line the sum is over primitive periodic orbits $p$
on the graph $\mathcal{G}_{II}$
and their $r$-th repetition. Note that the periodic orbits on $\mathcal{G}_{II}$
here differ from
the periodic orbits on $\mathcal{G}_I$ in the first approach which contains loops at
the vertices. A periodic orbit
$p=\overline{v_1v_2\dots v_{n_p}}$  of length $n_p \in \NN$ on $\mathcal{G}_{II}$  
consist of a sequence of $n_p$ vertices such that $v_{k}$ and $v_{k+1}$ are always
different and connected by an edge on $\mathcal{G}_{II}$ (we again set $v_0=v_{n_p}$
and $v_{n_p+1}=v_1$).
The weight of the periodic orbit may be written as
\begin{equation}
  {  \mathcal{W}_{II, p}}=
  \prod_{k=1}^{n_p} \sigma^{(v_k)}_{v_{k+1}, v_{k-1}}\ .
\end{equation}
The expression \eqref{countingfunction_periodicorbitsum}
contains an infinite
sum over periodic orbits of arbitrary length. Let us show, by
reference to well known facts, that all relevant information can be drawn
from relatively short orbits. For this we introduce the notion of a
pseudo-orbit $\mathcal{P}= \prod_{k=1}^M p_k^{m_k}$ which is just a formal
product of a finite set of periodic orbits $\{ p_k\}_{k=1}^M$ with
multiplicities $m_k$. We associate the length $n_{\mathcal{P}}=
\sum_{k=1}^M m_k n_{p_k}$
(where $n_{p_k}$ is the length of the periodic orbit $p_k$)
and the amplitude
\begin{equation}
  {  \mathcal{W}_{II,\mathcal{P}}}= \prod_{k=1}^M
  {  \mathcal{W}_{II, p_k}^{m_k}}\ .
\end{equation}
The main observation \cite{subdeterminant}
here is that the amplitude of a long periodic orbit
$p_0$
that travels through the same directed edges more than once can be expressed
as the amplitude of a pseudo-orbit $\mathcal{P}=p_1^{m_1}\dots p_M^{m_M}$
(with length $n_{\mathcal{P}}=n_{p_0}$) where none of the periodic orbits
$p_k$ visits any directed edge more than once. If this decomposition is
non-trivial (if $\mathcal{P}\neq p_0$) one calls $p_0$ reducible
otherwise (if $\mathcal{P}= p_0$) one calls $p_0$ irreducible.
Alternatively a periodic orbit is reducible if any directed edge is
visited more than once and otherwise irreducible. As irreducible orbits
have at most length $2E$ (the number of directed edges) only a finite number
of amplitudes contains all relevant spectral information.\\
Unitarity
of the discrete-time quantum evolution
operator $S_{II}(\lambda)$ implies the functional equation
\begin{equation}
  \zeta_{II}(\lambda)= \det\!\left(S_{II}(\lambda) \right)\
  \zeta_{II}(\lambda)^* = \zeta_{II}(\lambda)^* \prod_{v=1}^N
  \frac{H_{vv}-\lambda+i \Gamma_v}{H_{vv}-\lambda-i\Gamma_v}\ .
\end{equation}
The latter leads to a further reduction
in the number of independent
amplitudes such that only irreducible orbits $p$ of length
$p \le E$ (rather than $2E$) are required.
We refer to the literature \cite{subdeterminant, pseudoorbit}
for the complete systematic development of the pseudo-orbit approach
to trace formulas.

\subsection{Associated discrete-time classical random walk on the
  underlying graph}

The evolution operator $S_{II}(\lambda)$ defines a discrete-time
quantum walk
on the directed edges of the graph $\mathcal{G}_{II}$.
We may associate to this a discrete-time classical random walk on the
directed edges. For this, one replaces the quantum amplitude $S_{II}(\lambda)_{e' e}$
to scatter from one
directed edge $e$ to another directed edge $e'$ by the absolute square
\begin{equation}
  M(\lambda)_{e' e}= \left|S_{II}(\lambda)_{e' e} \right|^{  2}\ .
\end{equation}
The unitarity of $S_{II}(\lambda)$ then implies that
\begin{equation}
  \sum_{e} M(\lambda)_{e'e}=1= \sum_{e'} M(\lambda)_{e' e}
\end{equation}
where the sums are over all directed edges.
In other words $M(\lambda)$ is a bi-stochastic matrix and defines a Markov process
on the directed edges that is consistent with the connectivity of the graph,
in short a random walk.
An analogous
association of a corresponding classical dynamics on quantum graphs (i.e. metric
graphs
with a self-adjoint Schr\"odinger operator) has been very useful in applications
of quantum graphs in quantum chaos \cite{review}.

{ 
  \subsection{Some example applications of the evolution operator
    $S_{II}(\lambda)$
    and the corresponding trace formula}
  
  The trace formula \eqref{eq:traceformula_II} may be applied to a number of
  open interesting problems. For instance one may use it to explore spectral
  properties of random-matrix ensembles  of sparse matrices with a
  given sparsity pattern defined through the adjacency matrix of
  the corresponding graph. The trace formula \eqref{eq:traceformula_II}
  then allows to describe the mean density of states and spectral correlations
  in terms of ensemble averaged weights of periodic walks on the graph.
  This type of approach has been very useful in the past and may lead to new
  insights in the present case. Such an exploration is beyond the scope of this
  manuscript. However we would like to explore shortly
  two other applications of the evolution operator $S_{II}(\lambda)$:
  the Anderson model in one dimension and a one-parameter family of quantum walks
  related to a given Hermitian matrix $H$.
  
  \subsubsection{Jacobi matrices and the one-dimensional Anderson model on a
    chain}\label{sec:Anderson}
  
  The matrix $H$ considered here is a $N\times N $
  Jacobi matrix with arbitrary diagonal entries
  $H_{vv}$ and $H_{v,v-1}=H_{v-1,v} =1, \ v=2, \cdots N-1$
  on the two adjacent diagonals.
  The corresponding graph is a finite chain consisting
  of $N$ vertices connected linearly,
  with $N-2$ vertices of degree two, and at the two ends the vertices are of
  a degree $1$.
  The corresponding scattering matrices on the internal vertices  $(1<v<N)$ are
  \begin{equation}
    \sigma^{(v)}(\lambda) =
    \frac{i}{H_{vv}-\lambda-2i}
    \begin{pmatrix}
      H_{vv}-\lambda & 2i \\
      2i & H_{vv}-\lambda
    \end{pmatrix}   \ \
    =\ \
    ie^{i\phi_v(\lambda)/2}\begin{pmatrix}
      \cos \left(\phi_v (\lambda)/2\right)& i
    \sin \left(\phi_v(\lambda)/2\right) \\
    i\sin\left( \phi_v(\lambda)/2\right) &
    \cos\left( \phi_v(\lambda)/2\right)
  \end{pmatrix}
  \label{eq:jacobi_vertex_scattering}
  \end{equation}
  where
  $\phi_v(\lambda) =  2\mathrm{arccot}\left (\frac{H_{vv}-\lambda}{2}\right )$.
  In the Anderson model the diagonal elements are
  independent identically distributed random
  variables. If one fixes the spectral parameter in the Anderson model
  then the phases $\phi_v$
  are also independent identically distributed variables with a
  probability law
  that depends on the spectral parameter.
  For instance, if the diagonal elements are distributed according to a
  Cauchy law
  $P(H_{vv})=\frac{2}{\pi\left(4+(H_{vv}-\mu )^2\right)} $ centered at $\mu$
  then it is easy to show that the phases $\phi_v$ are uniformly distributed
  on the unit circle for $\lambda = \mu$.
  %%%%% 
  % (It is easy to show that if the diagonal matrix elements are iid random
  % variables, and  $\xi_k  =\frac{H_{kk}-\lambda}{2}$ distribute according
  % to the Cauchy distribution
  % $p(\xi)=\frac{1}{2\pi} \frac {1}{1+\xi^2}$, then
  % the corresponding phases $\phi_k$ are
  %uniformly distributed on the unit circle).
  
  At the end points $v=1,N$ the vertex matrix element is a phase
  \begin{equation}
    \sigma^{(1)}= i\frac{H_{11}-\lambda+i}{H_{11}-\lambda-i}= e^{i\phi_1(\lambda)/2}\  ;\  \sigma^{(N)}= i\frac{H_{NN}-\lambda+i}{H_{NN}-\lambda-i}= e^{i\phi_N(\lambda)/2} \ .
  \end{equation}
  
  Per definition, the vertex scattering matrix $\sigma ^{(v)}$ provides the
  linear relation between the outgoing amplitudes from the vertex $v$,
  $(a_{v-1,v},a_{v+1,v})$ and the incoming
  amplitudes $(a_{v,v+1},a_{v,v_1})$. It can be used in order to compute the
  % transmission
  transfer matrix $\tau^{(v)}$ which expresses the amplitudes
  pertaining to the edge $(v,v-1)$
  and $(v+1,v)$ such that
  \begin{equation}
    \begin{pmatrix}
      a_{v+1,v}\\
      a_{v,v+1}
    \end{pmatrix}
    =
    \tau^{(v)}
    \begin{pmatrix}
      a_{v,v-1}\\
      a_{v-1,v}
    \end{pmatrix}\ .
  \end{equation}
  %: $(a_{v,v-1},a_{v-1,v})$ to the amplitudes
  %pertaining to the next edge $(v+1,v)$ : $(a_{v+1,v},a_{v,v+1})$.
  A short calculation shows
  \begin{equation}
    \tau^{(v)}(\lambda) =
    -i
    \begin{pmatrix} 1 & 0\\
      0 & -1
    \end{pmatrix}
    -\cot\left(\phi_v(\lambda)/2\right)
    \begin{pmatrix}
      1 & i\\
      -i & 1
    \end{pmatrix}\ .     
    %\begin{pmatrix}
    %  0 &\  1 \\
    %  1 & \ 0
    %\end{pmatrix}   \
    %+\
    %\cot\phi_v(\lambda) \begin{pmatrix}
    %  1& -i \\ i & \ 1
    %\end{pmatrix}\ .
  \end{equation}
  The transfer matrices enable writing a secular equation for the spectrum
  of $H$: Since the eigenvectors are determined up to a constant scaling,
  choose $a_{1,2} =1$. Then, $a_{2,1} = e^{i\phi_1(\lambda)/2}$. Multiplying the
  transfer matrices along the chain gives 
  \begin{equation}
    \begin{pmatrix}
      a_{N,N-1}(\lambda)   \\
      a_{N-1,N}(\lambda) 
    \end{pmatrix}    
    =\ \prod_{k=2}^{N-1}\tau^{(k)}(\lambda)
    \begin{pmatrix}
      e^{i\phi_1(\lambda)/2} \\ \  1
    \end{pmatrix} \ .
  \end {equation}
  However, the ratio between $a_{N,N-1}(\lambda)$ and
  $a_{N-1,N}(\lambda)$ is $e^{-i\phi_N(\lambda)/2}$.
  This provides an equation which determines  exactly $N$ values of
  $\lambda$ which are the desired spectrum.
  The last statement can be proved by noticing that the matrix elements
  of $\tau(\lambda)$ are linear in $\lambda$ so their products are polynomials.

  \subsubsection{The one-parameter family of
    quantum walks and random walks associated to a
    Hermitian matrix $H$}
  \label{quantum_walks}

  The example of Jacobi matrices and the related Anderson model
  in the previous section is interesting in its own right.
  It also provides the a basic example for
  the link to the topic
  of
  quantum walks \cite{kempe}.
  This is a topic that we now want to explain in some detail
  more generally. Quantum walks are  often 
  discussed in connection with quantum search algorithms
  in the theory of
  quantum computation but are also interesting in their own
  right as a quantum version of the classical random
  walks.}\footnote{  Quantum walks
    have often been called `quantum random walks'
    because they are a quantum version of a classically 
    stochastic process.
    %However there is nothing random
    %in the quantum version -- indeed
    %randomness is being replaced by quantum features. We
    %believe that the name
    %`quantum random walks' is confusing and prefer just
    %`quantum walks.
    However there is nothing random in the quantum version – rather,
    `randomness' is an intrinsic feature of quantum (wave) dynamics .
    We believe that the name `quantum
    random walks’ is inappropriate and prefer just `quantum walks'.
  }
  { 
  Starting with an arbitrary Hermitian matrix $H$ we
  have defined a naturally related one-parameter family
  of unitary matrices
  $S_{II}(\lambda)$ that may be viewed as  discrete time
  quantum walks on the
  directed edges of a graph.
  Moreover, for each real $\lambda$ the latter may be
  viewed as a quantum version
  of a classical discrete time random walk on the
  directed edges defined by the matrix $M(\lambda)$.

  %Before going into the example of Jacobi matrices
  %on a chain graph let us
  %make a few general remarks about this connection.
  Discrete time
  quantum walks are often described as a walk on the
  vertices of a graph with an additional degree of freedom
  known as coin states which carry information about the
  direction (the
  previously visited vertex) \cite{kempe}. The difference
  between the standard formulation on vertices with coin states
  on one side and a
  formulation using directed edges 
  is superficial as
  there is a one-to-one correspondence. This one-to-one
  correspondence carries
  over to the definition of the evolution operator for a
  unit time step. In the standard formulation this is a
  product of two steps. One first chooses a new direction
  using a unitary coin operator and this is followed by a
  unitary shift operator which moves the quantum walker
  to the next vertex. It is easy to see that this is
  equivalent to the fact that $S_{II}(\lambda)$ is a product
  where one first acts with the unitary vertex
  scattering matrices (the coin operation) followed by a
  permutation that interchanges the direction on a
  given edge (the shift operation). 

  In the random walk the main object of interest is the probability
  $P_{vw}^{(\mathrm{RW})}(n)$ to find a walker on the directed edge from
  vertex $w$ to $v$ after $n$
  time steps. Combining these probabilities into a column vector
  $\mathbf{P}^{(\mathrm{RW})}(n)$
  the time evolution is defined iteratively by
  \begin{equation}
    \mathbf{P}^{(\mathrm{RW})}(n+1)=M(\lambda) \mathbf{P}^{(\mathrm{RW})}(n)\ .
  \end{equation}
  Assuming that the walker starts on one directed edge $v_0w_0$
  one may set $ P_{v_0w_0}^{(\mathrm{RW})}(0)=1$ and all other initial
  probabilities to zero.
  One then finds
  that the probability to find the walker on the directed edge $vw$ after
  $n$ steps may be written as a sum 
  \begin{equation}
    P_{vw}^{(\mathrm{RW})}(n) = \sum_p \mathcal{M}_p
  \end{equation}
  over all walks $p$ on the graph of length $n$ that start at $v_0w_0$
  and end at $vw$ with weights $\mathcal{M}_p$ that are products of the
  corresponding matrix elements of $M(\lambda)$.

  The quantum walk is defined analogously where the main object is now a
  set of quantum amplitudes $a_{vw}(n)$ that satisfy
  $\sum_{vw} \left|a_{vw} \right|^2=1$ (a sum over all directed edges of the
  underlying graph) with a time evolution
  \begin{equation}
    \mathbf{a}(n+1)= S_{II}(\lambda) \mathbf{a}(n)\ .
  \end{equation}
  We would like to stress that the quantum evolution of the quantum walk
  is fundamentally different to the proper time quantum  evolution
  $e^{-iHt}$.
  The latter \emph{cannot} be reconstructed in some way from the quantum walk.
  The probability to find the quantum walker on the directed edge $vw$
  after $n$ steps is given by
  \begin{equation}
    P_{vw}^{(\mathrm{QW})}(n)= \left|a_{vw}(n) \right|^2\ .
  \end{equation}
  If the quantum walker starts on the directed edge $v_0w_0$ we may set the
  corresponding amplitude $a_{v_0w_0}(0)=1$ and all other initial amplitudes
  to zero. The amplitude at the directed edge after $n$ steps is then again a
  sum
  \begin{equation}
    a_{vw}(n) = \sum_p \mathcal{W}_p
  \end{equation}
  over all walks from $v_0w_0$ to $wv$ of length $n$ where the weight
  $\mathcal{W}_p$ is the product of all quantum scattering amplitudes along
  the walk. Note that $\left|\mathcal{W}_p\right|^2= \mathcal{M}_p$.
  The probability at the directed edge $vw$ after $n$ steps is then
  a double sum over walks that may be written as 
  \begin{equation}
    \begin{split}
      P_{vw}^{(\mathrm{QW})}(n)=& \sum_{p,p'} \mathcal{W}_{p'}^* \mathcal{W}_p\\
      =& P_{vw}^{(\mathrm{RW})}(n) + \sum_{p\neq p'}
      \mathcal{W}_{p'}^* \mathcal{W}_p\ .
    \end{split}
  \end{equation}
  In the second line we have combined the diagonal part of the double
  sum to the corresponding probability of the corresponding classical random
  walk. This shows that any distincitive quantum effects are taking place in
  the
  off-diagonal sum over unequal pairs of walks $p \neq p'$.
  It is a well-known fact that the dynamic behaviour of the two probability
  distributions
  may be fundamentally different. For instance, letting the
  dimension of the Jacobi matrices in the previous section go to infinity any
  such Jacobi matrix defines a one-parameter family of random and quantum
  walks
  on the line. The long-time behaviour of the random walk on the line is
  generically
  diffusive: asymptotically for $n \to \infty$ the variance of the position of
  the walker increases proportional to $n$. For the quantum walk the long-time
  behaviour may differ strongly. In the presence of disorder Anderson
  localization sets in which inhibits any growth of the variance. On the other
  side in the absence of disorder it is well known that quantum walks on the
  line
  may behave ballistically which means that the variance grows proportional to $n^2$.
  Performing the corresponding sums over walks explicitly is a non-trivial
  combinatorial 
  task even for the line. In the absence of disorder it may be circumvented by
  explicit spectral decomposition. In the presence of disorder on the line 
  the combinatorics for the return probability has been solved in terms of
  recursion formulas \cite{Uzy_Holger}. We believe that similar combinatorial
  methods
  can lead to interesting new insights on these models. We will not pursue
  this further here as it goes clearly beyond the aim of this manuscript.
}

\section{Conclusion}
\label{conclusions}

In this work we have presented two rather different trace formulas that may be
used to analyse the spectrum of a Hermitian matrix $H$. Both approaches
have in common that the spectral information is contained in a unitary
matrix $S(\lambda)$ such that $\lambda$ is an eigenvalue of $H$ if and only
if $S(\lambda)$ has a unit eigenvalue.\\
In the first method we have made a connection to Polylogarithms
and Eulerian polynomials. The coefficients of these polynomials
are the Eulerian numbers which are of combinatorial character.
The relation to combinatorics was further explored in {  App.~\ref{appendix}}
where $\tr H^m$ for large $m>N$ is expressed in terms the first
$N$ traces using Newton identities and properties of the Eulerian numbers.\\
In the second approach the unitary matrix $S_{II}(\lambda)$ expresses the
spectral information of $H$ in a quantum random walk
on the directed edges of an associated graph such that each
non-vanishing off-diagonal matrix element of $H$ is represented by an edge.
The random walk depends
parametrically on the spectral parameter $\lambda$. We derive a trace formula
which may be expressed as a sum over periodic orbits of this quantum
random walk. It is interesting that the Gershgorin radius which gives
bounds on the spectrum is explicitly present in the trace formula
and the smooth part of the trace formula incorporates these bounds
in a smooth way: the smooth part of the density of states is a sum
over Lorentzians with positions and widths determined by Gershgorin's
theorem.

Apart from giving new interesting connections to combinatorics and
spectral theory both approaches give a new tool to understand spectra
of families of Hermitian matrices in terms of the underlying graph.
Among the potential future applications of the trace formulas are random matrix
ensembles for a fixed underlying graph -- e.g., given a connected
graph $\mathcal{G}$
with adjacency matrix $A_{vw}$ one may consider random matrices $H$
with off-diagonal matrix elements $H_{vw}=A_{vw} X_{vw}$ where $X_{vw}$
are random variables
distributed to some given law. For instance one may define relatives
of the well-known Gaussian ensembles GOE and GUE by choosing $X$ from
either of these ensembles. For large well connected graphs one then expects
to recover GOE or GUE behaviour while less well connected (sparse graphs, or
graphs with few bridges between large well connected subgraphs) will
show deviations. Expressing spectral information in terms of periodic orbits
via the  trace formulas that  we have presented here is a new potentially
fruitful tool for
understanding such random-matrix ensembles for a given graph.

\acknowledgments
We would like to thank Professor M.V.~Berry
for useful remarks concerning the material presented in Sec.~\ref{tf1},
and Professor P.~Deift for bringing the Gershgorin theorem to our attention. 

\appendix

\section{A few remarks on the numerical convergence of the trace formulas}
\label{appendix_convergence}

\begin{figure}
  \includegraphics[width=0.8\textwidth]{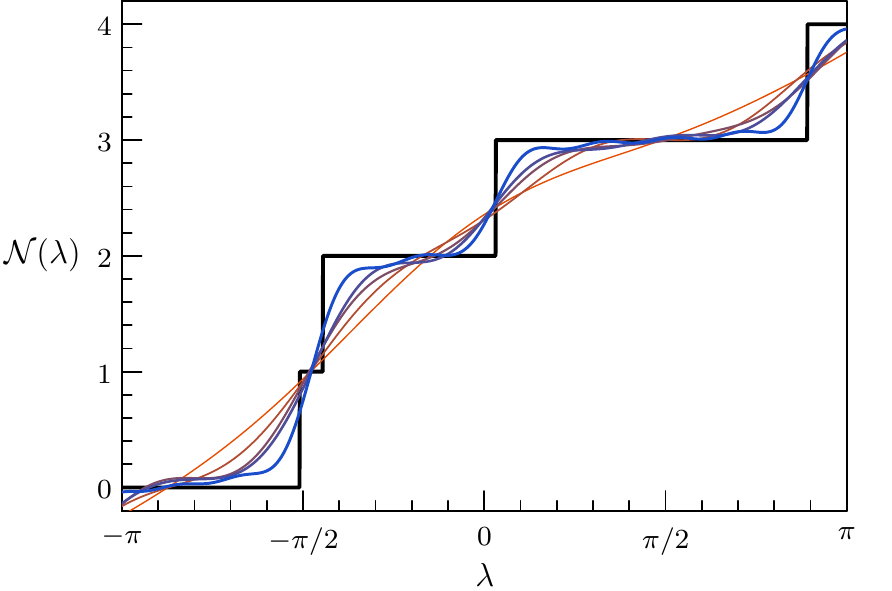}
  \caption{\label{fig} Spectral counting function $\mathcal{N}(\lambda)$
    for  a Hermitian $4\times 4$ matrix $H$ with
    eigenvalues $\lambda_1=-1.6$, $\lambda_2=-1.4$, $\lambda_3=0.1$, and $\lambda_4=2.8$ (thick line) together with approximations
    at
    finite resolution $\epsilon>0$ based on the trace formula \eqref{tracepoly}.
    The five approximations correspond to a cut-off $n_\mathrm{max}$
    taking values in $\{2,3,4,5,10\}$,  $s_{\mathrm{max}}=9 n_{\mathrm{max}}$
    and resolution $\epsilon=1/n_{\mathrm{max}}$.
  }
\end{figure}

As stated in the main text the
trace formulas \eqref{tracepoly} and \eqref{tracepoly1}
do not converge as an identity of functions but only in a weaker
sense of distributions acting on a suitable set of sufficiently smooth
test functions. This remains the case even if
one does not perform the implied limit and
keeps $\epsilon>0$. Absolute convergence is
only recovered for $\epsilon > \pi$ where any resolution of the spectrum
is lost.
In this appendix we want to consider how to make use of these trace formulas
if one has access to a finite number of traces $\tr\!\left(H^s\right)$
and we want to use the trace formula \eqref{tracepoly} 
to get some information about the location
of the spectrum.
Even at finite $\epsilon>0$ the expression \eqref{tracepoly1} in terms
of Polylogarithmic functions cannot be
used in this case (unless $\epsilon>\pi$) rather one needs to go back
to the double sum in the first line of \eqref{tracepoly_osc} where one first
sums over $s$ and then over $n$. This double sum converges
(though not absolutely) at finite $\epsilon>0$ to a sum of
step functions where the steps are smeared out over an interval of size
$\propto \epsilon$. For a reasonable approximation of the counting function at
a given resolution it is sufficient (and practical) to
introduce cut-offs
$n_\mathrm{max}$ and $s_\mathrm{max}$
such that only the traces for $s \le s_\mathrm{max}$ contribute.
These cut-offs depend on the resolution $\epsilon>0$
that one wants to achieve, and both cut-offs increase
{  without bound}
%unboundedly
($s_{\mathrm{max}}\to \infty$ and $n_{\mathrm{max}}\to \infty$)
as $\epsilon \to 0$. 
In order to estimate a reasonable choice of cut-offs
that ensures convergence to the exact counting function as $\epsilon \to 0$
one should first consider the summation over $s$ which is of exponential type.
The exponential series $e^z=\sum_{s=0}^\infty \frac{z^s}{s!}$
starts to converge when
$|z|^s < s!$. Using Stirling's formula a reasonable cut-off for the
exponential function is $s_\mathrm{max} > e |z|$. In our case
one should choose $|z|= \pi n_{\mathrm{max}}$ (as the eigenvalue spectrum
is bounded $-\pi <\lambda_k<\pi$). With $\pi e <9$ one may then choose 
$
  s_\mathrm{max} > 9 n_{\mathrm{max}}$.
Note that this cut-off depends on $n_{\mathrm{max}}$ rather than the resolution
$\epsilon$.
Once the exponential has converged the remaining sum over $n$
is of logarithmic type
$-\log(1-e^{-\epsilon})=\sum_{n=1}^\infty \frac{e^{-n\epsilon}}{n}$
and thus starts to converge when $n> 1/\epsilon$ which leads to the cut-off
$
  n_{\mathrm{max}}> \left[ \frac{1}{\epsilon}\right]
$
where $\left[x\right]$ denotes the smallest integer {  larger}
than (or equal to)
$x$. Altogether we find
\begin{equation}
  s_\mathrm{max} > 9\, n_{\mathrm{max}}> 9 \left[\frac{1}{\epsilon}\right] \ .
  \label{cut-off}
\end{equation}
At any finite $\epsilon>0$ one may increase the cut-offs and the sums converge
as long as the first inequality in \eqref{cut-off} is kept.\\
In Figure~\ref{fig} we show how this works in practice if $H$
is a matrix of dimension $N=4$ with eigenvalues
$\lambda_1=-1.6$, $\lambda_2=-1.4$, $\lambda_3=0.1$, and $\lambda_4=2.8$.

\section{Some relations following from the Worpitzky identity}
\label{appendix}

The Worpitzky identity \eqref{worpitzky}
can be used to derive many identities involving the Eulerian numbers.
Below we derive a few and use them to prove
the identity \eqref{identity3}.

Let us start by substituting
$z= \pm r$ with $r\in \mathbb{Z}^+$ and $r<s$ in the Worpitzky identity
\eqref{worpitzky}.
We get two sets of identities by removing all the terms in the product
which include a vanishing factor. This gives
\begin{eqnarray}
  \sum_{k=s-r}^{s-1}A(s,k)
  \left (
  \begin{array}{c}
    k+r\\
    s
  \end{array}
  \right )
  =r^s \ ,
\end{eqnarray}
and
\begin{eqnarray}
  \sum_{k=0}^{r-1}A(s,k)
  \left (
  \begin{array}{c}
    r+s-k-1\\
    s
  \end{array}
  \right )
  =r^s \ .
\end{eqnarray}

Other identities are obtained by first writing the binomial coefficients
in \eqref{worpitzky} as a polynomial of degree $s$ in the variable $z$
\begin{equation}
  \binom{z+k}{s}=
  \prod_{l=1}^s(z+k+1-l) = \sum_{j=0}^s \alpha_j (s,k)  z^j
\end{equation}
where $\alpha_j(s,k)$ are the coefficients of this polynomial.
Then  \eqref{worpitzky} reads
\begin{equation}
  \sum_{j=0}^s z^j\, \left[\sum_{k=0}^{s-1} A(s,k)\alpha_j(s,k)\right] =s!z^s
\end{equation}
Hence, all the coefficients in the polynomial on the left hand side
must vanish, except for the $j=s$ one for which, using
\eqref{sum_Eulerian_numbers} one finds
$\alpha_s(s,k)=1$.\\
Denoting $t_r = \sum_{q=1}^{s}(k+1-q)^r$,  the Newton identities
enable one to write
\begin{equation}
  \label{newteq}
  \alpha_{s-m}{(s,k)} =\frac{1}{m!}
  \left|
    \begin{array}{ccccc}
      t_1 & 1   & 0 & \cdots & 0 \\
      t_2 & t_1 & 2 & \cdots & 0 \\
      \vdots & \vdots & \ddots &  \ddots & \vdots\\
      t_{m-1} &  t_{m-2} & \cdots  & t_1& m-1\\
      t_{m} & t_{m-1} & \cdots & t_2 & t_{1}
    \end{array}
  \right|.
\end{equation}
The terms for $j=0$ vanishes identically since
$\alpha_0(s,k)=\prod_{m=1}^s (k+1-m)=0 \ \ \forall \  0\le k\le s-1 $.
The next simple identities are obtained for  $j=1 \ {\rm and} \ j=s-1$ :
\begin{subequations}
  \begin{align}
    j= 1 : \qquad\quad&
              \sum_{k=0}^{s-1}
              A(s,k)\left(
              \prod_{m \ne k+1}^s
              (k+1-m)
              \right )
              &&=\nonumber \\
           &
             \sum_{k=0}^{s-1} A(s,k) (-1)^{s-1-k} k!(s-1-k)!&&= 0\ ,\\
    \qquad j=s-1  :\qquad\quad&
              \sum_{k=0}^{s-1} A(s,k)\left ((k+1)s-\frac{s(s+1)}{2}\right)
              &&= 0 \ .\qquad \qquad\quad
  \end{align}
\end{subequations}
Further identities can be written by expressing the $\alpha_j{(s,k)}$
explicitly.

In the rest of this appendix we derive the identity \eqref{identity3}.
Let us start by introducing
\begin{equation}
  B^{(m)}(s,k)
  =
  \left [\frac{ d^s }{dz^s}
    \left ( z^k(\log z)^m\right )  \right ]_{z=1}
  \label{Bdefinition}
\end{equation}
as a short-hand for the expression in the square brackets in identity
\eqref{identity3}.
This can
be computed to give
\begin{align}
  B^{(m)}(s,k)
  =
  &
    s! (-1)^m
    \sum_{l=\mathrm{max}(s-k,m)}^s
    (-1)^l
    \begin{pmatrix} 
      k \\
      s-l
    \end {pmatrix}
  \sum _{\underline{j}\in \mathcal{P}(k,l)}
  \left (\prod_{i=1}^k j_i \right )^{-1}
  \ ,
  \label{Bcoef}
\end{align}
where, $\mathcal{P}(k,l)$ is the set of partitions
of $l$ to $k$ integers $(j_1,j_2,\cdots,j_k)  \doteq \underline{j}$,
and all the $j_i$ are positive definite, with $\sum_{i=1}^k j_i=l$.
It is easy to show that $B^{m}(m,k)= m! \forall \ m-1 \ge k \ge 0 $,
which together with
\eqref{sum_Eulerian_numbers}
satisfies \eqref{identity3} for $s=m$.
Next let us prove the identity
\begin{equation}
  B^{(m)}(s,k)= \frac{\alpha_m(s,k)}{m!}
  \label{identity4}
\end{equation}
which will lead directly to \eqref{identity3}.
To show \eqref{identity4}
substitute
$ z=e^x$ in \eqref{Bdefinition}
so that
\begin{equation}
  B^{(m)}(s,k)= \left [(e^{-x} \frac{d}{dx})^s e^{kx} x^m\right ]_{x=0}\ .
  \label{Bdefinition2}
\end{equation}
Next define the generating function
\begin{equation}
  F(s,k;y)=\sum_{m=0}^\infty B^{(m)}(s,k) \frac{y^m}{m!}\ ,
\end{equation}
substitute \eqref{Bdefinition2} and
sum to obtain
\begin{equation}
  F(s,k;y)=\binom{k+y}{s} 
\end{equation}
which is exactly the polynomial whose coefficients are the
$\alpha_j(s,k)$ as defined above. This proves \eqref{identity3}.

\section{Examples}
\label{appendix_examples}

For illustration we work out explicitly the
discrete-time quantum evolution operator $S_{II}(\lambda)$ and
the corresponding spectral determinant $\zeta_{II}(\lambda,z)$
if the underlying graph is either an interval (the complete graph with
$N=2$ vertices) or a 2-star (equivalently, a chain of $N=3$ vertices).

\subsection{The interval ($N=2$)}

For a Hermitian $2\times 2$ matrix  $H$ with $H_{12}\neq 0$
the corresponding graph is just a single interval and the quantum evolution
operator takes
the form
\begin{equation}
  S_{II}(\lambda)=
  \begin{pmatrix}
    0  & \sigma^{(2)}_{11}(\lambda)\\
    \sigma^{(1)}_{22}(\lambda) & 0
  \end{pmatrix}
\end{equation}
where $\sigma^{(1)}_{22}(\lambda)$ and $\sigma^{(2)}_{11}(\lambda)$ are
unimodular
scattering phases
\begin{align}
  \sigma^{(1)}_{22}(\lambda)
  &
    =i \frac{H_{11}-\lambda +i h_{12}}{H_{11}-\lambda - i h_{12}}
    =i \frac{(H_{11}-\lambda)^2 - |H_{12}|^2 +2i h_{12}(H_{11}-\lambda)}{
    (H_{11}-\lambda)^2 + |H_{12}|^2}
  \\
  \sigma^{(2)}_{11}(\lambda)
  &
    = i \frac{H_{22}-\lambda +i h_{12}}{H_{22}-\lambda - i h_{12}}
     =i \frac{(H_{22}-\lambda)^2 - |H_{12}|^2 +2i h_{12}(H_{22}-\lambda)}{
    (H_{22}-\lambda)^2 + |H_{12}|^2}
    \ .
\end{align}
The determinant is given by
\begin{equation}
  \begin{split}
    \det\left( S_{II}(\lambda) \right) =&
    -\sigma^{(1)}_{22}(\lambda) \sigma^{(2)}_{11}(\lambda)\\
    =&
    \frac{\zeta_H(\lambda) +i h_{12}\left(H_{11}+H_{22}-2\lambda
      \right)}{
      \zeta_H(\lambda) -i h_{12}\left(H_{11}+H_{22}-2\lambda
      \right)
    }
  \end{split}
\end{equation}
and the spectral determinant
is
\begin{equation}
  \zeta_{II}(\lambda,z)=
  1-z^2\sigma^{(1)}_{22}(\lambda) \sigma^{(2)}_{11}(\lambda)
  =\left(1- z\sqrt{\sigma^{(1)}(\lambda)\sigma^{(2)}(\lambda)}\right)
  \left(1+z \sqrt{\sigma^{(1)}(\lambda)\sigma^{(2)}(\lambda)}\right)
\end{equation}
which is consistent with the fact that
the two eigenvalues of $S_{II}(\lambda)$ (which can be read directly from the
matrix itself) are
\begin{equation}
  \begin{split}
    z_1=e^{i\theta_1(\lambda)}=& \sqrt{\sigma^{(1)}_{22}(\lambda)
      \sigma^{(2)}_{11}(\lambda)}\\
    z_2=e^{i\theta_2(\lambda)}=& -\sqrt{\sigma^{(1)}_{22}(\lambda) \sigma^{(2)}_{11}(\lambda)}
  \end{split}\ .
\end{equation}
Rewriting the spectral determinant as
\begin{equation}
  \zeta_{II}(\lambda,z)=1-z^2+z^2 \zeta_{II}(\lambda)
\end{equation}
with
\begin{equation}
  \zeta_{II}(\lambda)=
  \frac{2 \zeta_H(\lambda)}{
    \left(H_{11}-\lambda -ih_{12}\right)
    \left(H_{22}-\lambda -i h_{12}\right)}
\end{equation}
shows that $\zeta_{II}(\lambda)$ has the same zeros as $\zeta_H(\lambda)$
in the complex $\lambda$-plane. 

\subsection{The two-star graph ($N=3$)}

Next consider the Hamiltonian of the form
\begin{equation}
	H=
	\begin{pmatrix}
		H_{11} & H_{12} & H_{13}\\
		H_{21} & H_{22} & 0\\
		H_{31} & 0 & H_{33}
	\end{pmatrix} .
\end{equation}
This corresponds to a two-star graph with the central vertex
$v=1$ of degree $d_1=2$
and two vertices $v=2$ and $v=3$ of degree one.
The spectral determinant of the Hamiltonian is then
\begin{equation}
	\zeta_H(\lambda)=
	\left(H_{11}-\lambda\right)\left(H_{22}-\lambda\right)\left(H_{33}-\lambda\right) 
	-\left(H_{22}-\lambda\right)|H_{13}|^2 - \left(H_{33}-\lambda\right)|H_{12}|^2\ .
\end{equation}
With $\Gamma_1= h_{12}+ h_{13}\equiv |H_{12}| + |H_{13}|$
the corresponding vertex scattering matrices are
\begin{equation}
  \begin{split}
    \sigma^{(1)}=&
    \frac{i}{H_{11}-\lambda-i\Gamma_1}
    \begin{pmatrix}
      \left(H_{11}-\lambda+i (h_{12}-h_{13})\right)
      & i2 \sqrt{h_{12}h_{13}}e^{i(\gamma_{13}-\gamma_{12})}\\
      i2 \sqrt{h_{12}h_{13}}e^{-i(\gamma_{13}-\gamma_{12})}&
      \left(H_{11}-\lambda+i (h_{13}-h_{12})\right)
    \end{pmatrix}\\
    \sigma^{(2)}\equiv&\sigma^{(2)}_{11}=
    i\frac{H_{22}-\lambda+ih_{12}}{H_{22}-\lambda-ih_{12}}\\
    \sigma^{(3)}\equiv&\sigma^{(3)}_{11}=
    i\frac{H_{33}-\lambda+ih_{13}}{H_{33}-\lambda-ih_{13}}\ .
  \end{split}
\end{equation}
The quantum evolution operator has the form 
\begin{equation}
  S_{II}(\lambda)=
  \begin{pmatrix}
    0 & 0 & \sigma^{(2)}_{11} & 0\\
    0 & 0 & 0 & \sigma^{(3)}_{11}\\
    \sigma^{(1)}_{22} & \sigma^{(1)}_{23} & 0 & 0\\
    \sigma^{(1)}_{32} & \sigma^{(1)}_{33} & 0 & 0
  \end{pmatrix}
\end{equation}
and has determinant
\begin{equation}
  \det\ S_{II}(\lambda) =
  \sigma^{(2)}_{11}\sigma^{(3)}_{11}\det\ \sigma^{(1)}
  =\prod_{v=1}^3 \frac{H_{vv}-\lambda+i \Gamma_v}{H_{vv}-\lambda-i\Gamma_v}
  = \frac{G(\lambda) +i F(\lambda)}{G(\lambda)-i F(\lambda)}
\end{equation}
where
\begin{equation}
  \begin{split}
    G(\lambda)=&\mathrm{Re}
    \left(\prod_{v=1}^3 (H_{vv}-\lambda + i \Gamma_v)\right)\\
    =&
    \zeta_H(\lambda)-\Gamma_2\Gamma_3 \mathrm{tr} ( H-\lambda)\\
    F(\lambda)=&\mathrm{Im}
    \left(\prod_{v=1}^3 (H_{vv}-\lambda + i \Gamma_v)\right)\\
    =&
    (H_{22}-\lambda)(H_{11}+H_{33}-2\lambda)\Gamma_{3}+
    (H_{33}-\lambda)(H_{11}+H_{22}-2\lambda)\Gamma_{2}-\Gamma_1\Gamma_{2}\Gamma_3
  \end{split}
\end{equation}     
The spectral determinant can be calculated by direct calculation as
\begin{equation}
  \begin{split}
    \zeta_{II}(\lambda,z)=
    &
    1+z^2
    \frac{2\zeta_H(\lambda) +2\Gamma_{2}\Gamma_{3} \mathrm{tr}
      \left(H-\lambda\right)}{(H_{11}-\lambda-i\Gamma_1)(H_{22}-
      \lambda-i\Gamma_2)(H_{33}
      -\lambda-i\Gamma_3)}
    +z^4\det\ S_{II}(\lambda)\\
    =&
    \frac{
      \zeta_H(\lambda) \left(1+z^2\right)^2 -\Gamma_{2}\Gamma_{3}\mathrm{tr}\left( H-\lambda\right) \left(1-z^2 \right)^2+
      i(z^4-1) F(\lambda)
    }{\prod_{v=1}^3 \left( H_{vv}-\lambda -i \Gamma_v\right)}\ .
  \end{split}
\end{equation}
At $z=1$ this reduces to
\begin{equation}
  \zeta_{II}(\lambda)=
  \frac{4
    \zeta_H(\lambda)
  }{\prod_{v=1}^3 \left( H_{vv}-\lambda -i \Gamma_v\right)}\ .
\end{equation}
Note that the spectral determinant $\zeta_{II}(\lambda,z)$ is
bi-quadratic and the
zeros can be calculated directly as
\begin{equation}
  \begin{split}
    z_n=&e^{i \theta_n(\lambda)}\\
    =&
    \pm i
    \sqrt{
      \frac{\zeta_H(\lambda) +\Gamma_{2}\Gamma_{3} \mathrm{tr}( H-\lambda)
        \pm i
        \sqrt{\prod_{v=1}^3 |H_{vv}-\lambda +i \Gamma_v|^2 -
          \left(\zeta_H(\lambda) +\Gamma_{2}\Gamma_{3} \mathrm{tr} (H-\lambda)\right)^2
        }
      }{
        \prod_{v=1}^3(H_{vv}-\lambda +i \Gamma_v)}
    }
  \end{split}
\end{equation}
where $n\in\{1,2,3,4\}$ refers to the 4 different choices of signs.

\end{document}